\documentstyle[epsfig,aps,prc,eqsecnum]{revtex}

\def\lixo#1{}

\begin{document}


\title{Relativistic $O(q^4)$ two-pion exchange\\
nucleon-nucleon potential:\\
configuration space}

\author{
R. Higa$^{1,}$\footnote{higa@if.usp.br}, 
M. R. Robilotta$^{1,}$\footnote{robilotta@if.usp.br} and  
C.A. da Rocha$^{2,}$\footnote{carocha@usjt.br}}

\address{${}^{1}$ Instituto de F\'{\i}sica, Universidade de S\~{a}o Paulo,\\
C.P. 66318, 05315-970, S\~ao Paulo, SP, Brazil}

\address{${}^{2}$ N\'ucleo de Pesquisa em Computa\c c\~ao e Engenharia, 
Universidade S\~ao Judas Tadeu,\\
Rua Taquari, 546, 03166-000, S\~ao Paulo, SP, Brazil}

\date{\today}

\maketitle
\abstract{We have recently performed a relativistic $O(q^4)$ 
chiral expansion of the two-pion exchange $NN$ potential, and here we 
explore its configuration space content. Interactions are determined 
by three families of diagrams, two of which involve just $g_A$ and 
$f_{\pi}$, whereas the third one depends on empirical coefficients fixed 
by subthreshold $\pi N$ data. In this sense, the calculation has no 
adjusted parameters and gives rise to predictions, which are tested 
against phenomenological potentials. The dynamical structure of the eight 
leading 
non-relativistic components of the interaction is investigated and, in 
most cases, found to be clearly dominated by a well defined class of diagrams. 
In particular, the central isovector and spin-orbit, spin-spin, and tensor 
isoscalar terms are almost completely fixed by just $g_A$ and
$f_{\pi}$. The convergence of the chiral series in powers of 
the ratio (pion mass/nucleon mass) 
is studied as a function of the internucleon distance and, for $r>$ 1 fm, 
found to be adequate for most components of the potential. An important 
exception is the dominant central isoscalar term, where the convergence 
is evident only for $r>$ 2.5 fm. Finally, we compare the spatial behavior 
of the functions that enter the relativistic and heavy baryon formulations 
of the interaction and find that, in the region of physical interest, 
they differ by about 5\%.}

\section{introduction} \label{secI}

The research programme for the study of nuclear interactions was 
outlined more than fifty years ago, in a seminal paper by Taketani, 
Nakamura, and Sasaki \cite{TNS}. Pions, then recently detected, were 
identified as the relevant degrees of freedom for the construction of 
a theoretical potential. One pion exchanges would dominate at large 
distances, the exchanges of two uncorrelated pions would come next, 
and a square well could be used to simulate short range processes. 
It is quite remarkable that these ideas could stand for such a long 
time, survive the QCD revolution, and still remain as the qualitative 
framework of contemporary research. On the other hand, when the $NN$ 
research programme was first established, no precise information concerning 
the intrinsic structure of pions and their interactions  with nucleons 
was available. It took about forty years of intense collective work, 
both experimental and theoretical, for this aspect of the problem 
to be tamed, with the formulation of chiral perturbation theory (ChPT). 

The present day rationale for describing nuclear interactions by means 
of chiral symmetry is that low energy processes are strongly dominated 
by the quarks $u$ and $d$ and one may work with a two-flavor QCD. As 
the masses of these quarks are small in the GeV scale, one treats them 
as perturbations in a massless Lagrangian. The theory is 
symmetric under the Poincar\'e group and, in this limit, 
also invariant under both 
isospin and chiral SU(2)$\times$ SU(2) transformations. This last 
symmetry is realized in the Nambu-Goldstone 
mode and the QCD vacuum contains a condensate, 
that allows collective excitations, identified as 
pions. The non-Abelian character of QCD prevents low energy perturbative 
calculations and, in practice, one works with chiral effective theories, 
in which point-like baryons interact by exchanging pions that have 
small masses. 

The one-pion exchange potential (OPEP)  became definitively established 
in the early 1960s and is assumed to dominate completely $NN$ partial 
waves with orbital angular momentum $L\geq 5$. Its mathematical form was 
determined in the 1950s and remains stable ever since. One has 
also learned that any $\pi N$ interaction Lagrangian, based on either 
pseudoscalar or pseudovector couplings, chiral symmetric or not, yields 
the very same OPEP. Chiral symmetry is thus irrelevant for this part 
of the force, as for all single pion processes. 

The very opposite happens with the next layer of the interaction, the 
two-pion exchange potential (TPEP). This component is closely related to 
the $\pi N$ scattering amplitude and chiral symmetry becomes extremely 
important. In the 1960s, no perturbative treatment for strong 
interactions was available \cite{PSI} and potentials were constructed 
which incorporated $\pi N$ information by means of dispersion 
relations \cite{P}. In the same decade, chiral symmetry was being developed 
in a different framework and, with the help of current algebra techniques, 
low energy theorems for many pionic amplitudes were derived. 
Applications of chiral symmetry to $NN$ interactions \cite{SB}, 
three-body forces \cite{TM}, and exchange currents \cite{RW} began to be 
performed in the 1970s. At the end of this decade, Weinberg 
\cite{W1} outlined a research programme based on the idea of ChPT. 
In the 1980s this theory was fully developed
for the meson sector \cite{GL} and began to be used in the study of 
meson-baryon interactions \cite{GSS}. 

The systematic use of ChPT in the study of nuclear 
forces began in the early 1990s, through the works of Weinberg \cite{WNN} 
and Ord\'o\~nez and van Kolck \cite{OvK}, followed by other authors 
\cite{ChNN,RR94}. These early attempts to construct a chiral TPEP 
considered only pion and nucleon degrees of freedom and gave rise to 
poor descriptions of $NN$ data. Realistic potentials require other 
degrees of freedom, which were introduced in the form of deltas 
\cite{DNN}, hidden within $\pi N$ subthreshold coefficients 
\cite{SNN,RR97}, or incorporated into low energy constants (LECs) of 
effective Lagrangians \cite{KBW,Nij,EGM,K3,K4}. In spite of apparent 
differences, there must be a rather important overlap among these various 
approaches. This is expected because the numerical values of the LECs 
are normally obtained from empirical $\pi N$ subthreshold coefficients 
which, in turn, are largely dominated by delta intermediate states 
\cite{H}. So, to a large extent, one is just using different languages 
to express the same physics. Support to this view comes from the fact 
that potentials based on deltas \cite{DNN}, subthreshold coefficients 
\cite{BRR}, or LECs \cite{KBW,Nij,EGM,EM1} 
could produce satisfactory descriptions of asymptotic $NN$ phase 
shifts, without free parameters. This 
suggests that, if one could control carefully the peculiarities of 
the various approaches, the hope of having a TPEP as unique as the 
OPEP could be realized. This uniqueness is of major theoretical 
importance, since it would indicate that the effective theory can 
indeed represent QCD. 

In ChPT one uses a typical scale $q$, set by either 
pion four-momenta or nucleon three-momenta, such that $q\ll 1$ GeV. 
The leading term of the chiral TPEP is $O(q^2)$ and, at 
present, there are two independent expansions of the potential up to 
$O(q^4)$ in the literature. The first one is based on heavy baryon 
chiral perturbation theory (HBChPT) \cite{KBW,K3}, where one uses 
non-relativistic Lagrangians from the very beginning and the inverse 
of the nucleon mass $(m)$ as an expansion parameter. Relativistic 
corrections, needed at $O(q^4)$, are added separately \cite{K4}. 
The alternative calculation was produced recently by ourselves \cite{HR}, 
is covariant, and results were expressed directly in terms of 
loop integrals and observable subthreshold $\pi N$ coefficients. In the 
case of $\pi N$ scattering, heavy baryon \cite{FMS} and relativistic 
\cite{BL1,BL2} results do not coincide, due to the presence of some 
diagrams \cite{EllisT,UGM} that cannot be represented by series in powers 
of $q/m$. The same class of diagrams is present in the TPEP and the 
relativistic potential also cannot be expanded in the heavy baryon series 
around the point $t=4{\mu}^2$. If this restriction is nevertheless ignored 
and the $q/m$ expansion is performed in the relativistic potential, one 
recovers most of the structure produced by the heavy baryon formalism. The 
main discrepancies take the form of both $O(q^3)$ 
and $O(q^4)$ terms, associated, respectively, with the iteration of 
the OPEP and the Goldberger-Treiman discrepancy. The latter could be easily 
incorporated into the heavy baryon formalism, whereas the former may 
derive from the particular definition adopted for the potential, that 
has to suit a given dynamical equation. In our calculation we treated 
the iterated OPEP as in Ref.\cite{PL}.

QCD is a well defined theory and the same should happen when one works 
at the effective level. In the case of the TPEP, we consider the 
partial convergence between heavy baryon and covariant results at 
$O(q^4)$ as a rather welcome indication that uniqueness may not be 
too far ahead. The considerable narrowing of the theoretical discussion 
in the last decade represents a measure of the progress promoted by the 
systematic use of chiral symmetry, which has allowed one to understand 
the internal hierarchies of the $NN$ potential in terms of chiral layers. 
Nevertheless, the question still remains open as to the extent 
this mathematical picture is backed by nature. 

The chiral picture may be assessed by comparing theoretical and 
empirical phase shifts. In the case of peripheral waves, this can be 
done perturbatively, without resorting to free parameters, and the 
main trends are well reproduced \cite{BRR,KBW,EM1}. However, these 
waves are small, error bars are important, and the test is not very 
stringent. In order to include inner waves, which are larger, 
one has to use dynamical 
equations, but this demands the regularization of the chiral potential 
by means of form factors, since it diverges at large momenta. The 
problem with form factors is that their use amounts to extending the 
dynamical content of the potential to higher orders in $q/m$ by means of 
free parameters, in a way which is not controlled by chiral symmetry. 
This blurs the chiral content of the potential and success in reproducing 
data, welcome as it is, also does not represent a very stringent test. 

As an interesting alternative, we may {\em assume} that the chiral 
potential, calculated at a given order, determines completely the 
interaction from a radius $R$ onwards and then use it as an input in phase 
shift analyses. 
This would just amount to extending to the TPEP a procedure which has 
already been used for a long time in the OPEP. For the latter, this 
idea has proved to be reliable in the elastic regime and for waves with 
$L\geq 5$. From the standpoint of the symmetry, this happens 
because chiral corrections are short ranged and one sees just the 
leading contribution through this window, irrespectively of the order 
in $q$ one is working at. In the case of the TPEP, the 
corresponding problem is much more complex and not fully understood. 
Works along this line have already been performed by the Nijmegen 
group \cite{Nij}, who claim that a $O(q^3)$ potential is effective 
for distances smaller than 2 fm. However, its conclusions are 
disputed by Entem and Machleidt \cite{EM-Nij} and the situation 
remains unclear. 

The present paper is motivated by the feeling that the quantitative 
aspects of chiral hierarchies need to be clearly understood if the 
TPEP is ever to become a reliable tool to be used in phase shift analyses. 
Our study is based on the configuration space version of the $O(q^4)$ 
potential produced in Ref.\cite{HR} and organized as follows. 
In Sec. \ref{secII} we 
discuss the dynamical content of the TPEP, which is given by a set 
of Feynman diagrams, organized into three families. 
The explicit expressions and corresponding figures for the various 
components of the potential are given in Sec. \ref{secIII}. 
As the way chiral symmetry is implemented 
varies with the family considered, in Sec. \ref{secIV} we discuss 
how dynamics is mapped into the final form of the potential and 
show that the importance of the LECs is rather channel dependent. 
Sec. \ref{secV} deals with the convergence of 
the chiral series and in Sec. \ref{secVI} we discuss the main 
differences between the relativistic and heavy baryon approaches to the 
potential. Finally, conclusions are presented in Sec. \ref{secVII}.

\section{dynamics} \label{secII}

\begin{figure}[!htb]
\begin{center}
\epsfig{figure=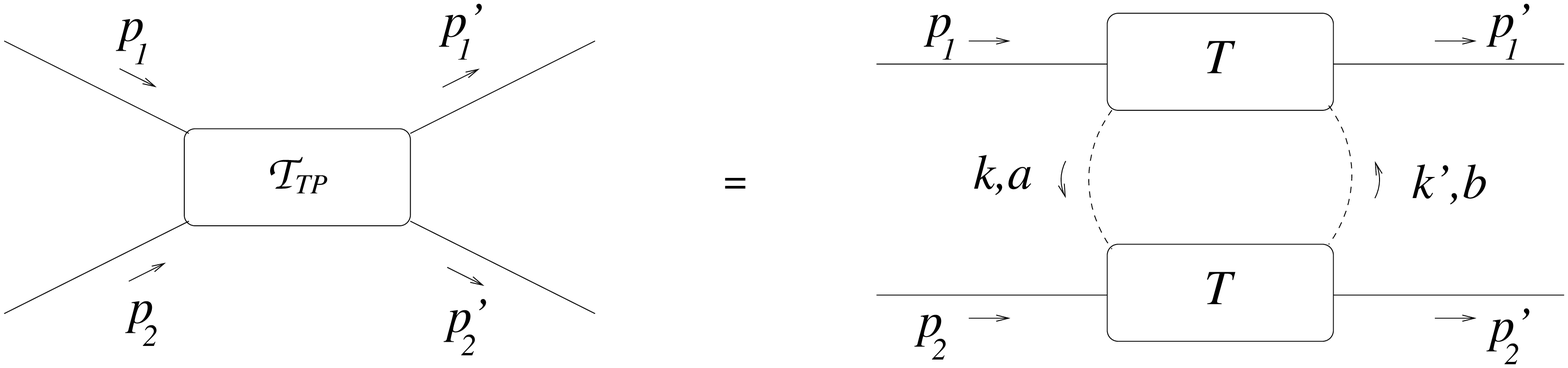, width=9cm}
\end{center}
\caption{Two pion exchange amplitude.}
\label{fig1}
\end{figure}

The dynamical content of the relativistic $O(q^4)$ chiral TPEP is 
determined by the effective Lagrangian 

\begin{equation}
{\cal L}_{eff}= {\cal L}_\pi^{(2)} + {\cal L}_N^{(1)}+{\cal L}_N^{(2)}
+{\cal L}_N^{(3)}\;,
\label{2.1}
\end{equation}  

\noindent
where ${\cal L}_\pi^{(n)}$ and ${\cal L}_N^{(n)}$ describe pion-pion and 
pion-nucleon interactions at $O(q^n)$. Other degrees of freedom are 
implicitly taken into account by means of the LECs $c_i$ and $d_i$, 
present in $ {\cal L}_N^{(2)}$ and ${\cal L}_N^{(3)}$. The use of covariant 
Feynman rules with vertices derived from this Lagrangian 
allows the construction of the $T$ matrix ${\cal T}_{T\!P}^{(4)}$, which 
describes the on-shell process $N(p_1)\;N(p_2)\rightarrow N(p'_1)\;N(p'_2)$ 
and contains two intermediate pions, as represented in Fig. \ref{fig1}. 
The potential is obtained by going to the center of mass frame and 
subtracting the iterated OPEP, in order to avoid double counting when it 
is used in the Lippmann-Schwinger equation. 

\begin{figure}[!htb]
\begin{center} 
\epsfig{figure=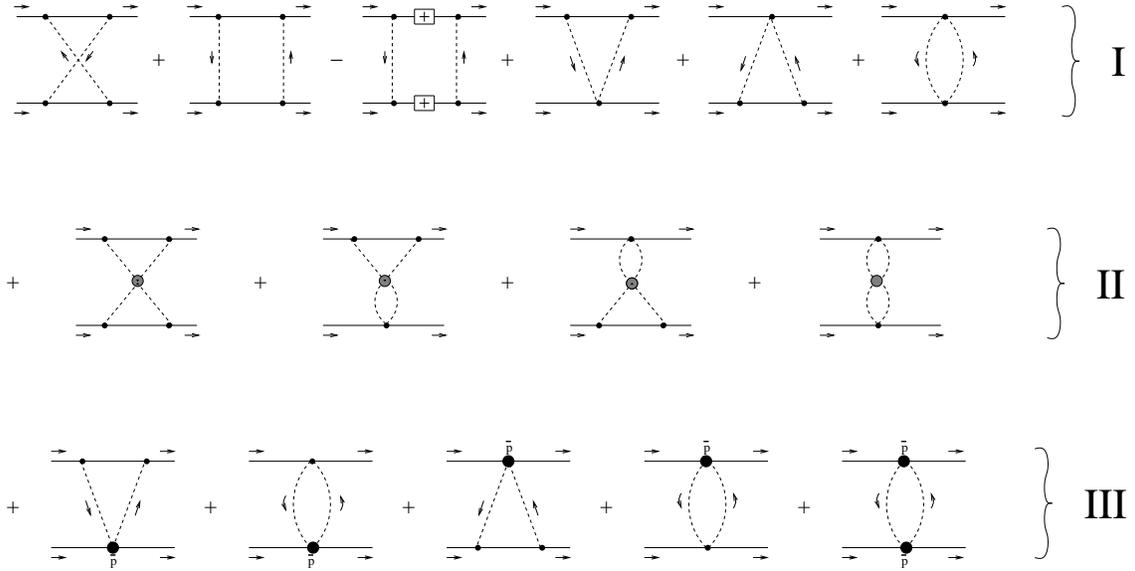, height=7.5cm} 
\end{center}
\caption
{Dynamical structure of the TPEP. The first two diagrams of family I 
correspond to the products of Born $\pi N$ amplitudes, the third one 
represents the iteration of the OPEP and the next three involve contact 
interactions associated with the Weinberg-Tomozawa vertex. The diagrams 
of family II describe medium range effects due to pion-pion correlations. 
Interactions represented by family III are {\em triangles} and 
{\em bubbles}, involving $\pi N$ subthreshold coefficients, indicated by 
the large black dots.}
\label{fig2}
\end{figure}

The dynamical content of the amplitude ${\cal T}_{T\!P}^{(4)}$ is given by 
the diagrams of Fig. \ref{fig2}. Their full evaluation produces amplitudes 
containing many different loop integrals, which are interconnected. The 
chiral orders of the potential are extracted by exploring as 
much as possible the mathematical relations among the various loop 
integrals. As the use of these results represents an important 
step in the determination of the potential, in the Appendix 
we display their accuracy in configuration space.

The processes given in Fig. \ref{fig2} are organized into three different 
families. The first one corresponds to the minimal realization of chiral 
symmetry \cite{RR94}, includes the subtraction of the iterated OPEP and 
involves only the pion-nucleon interactions given by $ {\cal L}_N^{(1)}$, 
with the constants $m$, $g$ and $f_\pi$ used at their physical 
values. The second family contains two-pion correlations in the 
$t$ channel, determined by ${\cal L}_N^{(1)}$ and 
${\cal L}_\pi^{(2)}$. Finally, the last family includes chiral corrections 
representing either higher order processes or other degrees of freedom, 
hidden into the LECs of ${\cal L}_N^{(2)}$ and ${\cal L}_N^{(3)}$.

This theoretical structure has been fully incorporated into our recent 
evaluation of the amplitude ${\cal T}_{T\!P}^{(4)}$, Ref.\cite{HR}. In that 
work we have performed a two-step calculation, using the fact that the 
$NN$ interaction is closely associated with the off-shell $\pi N$ 
amplitude. This allows one to use many 
of the results derived by Becher and Leutwyler \cite{BL2} (BL) for the 
$\pi N$ amplitude as inputs into the evaluation of the $NN$ potential. 
Moreover, it clarifies the relationship between the chiral orders of 
the $NN$ and $\pi N$ amplitudes. Using Fig. \ref{fig1}, we write the 
$O(q^n)$ expansion of ${\cal T}_{T\!P}$ as 

\begin{equation}
{\cal T}_{T\!P}^{(n)} = -\;\frac{i}{2!}\;\frac{1}{(2\pi)^4} \int \left\{ 
\frac{d^4Q}{[k^2-\mu^2]\;[k'^2-\mu^2]} \right\} \; 
\sum_{l,m}^{l+m=4}
[T_{\pi N}^{(l)}]^{(1)}\;[T_{\pi N}^{(m)}]^{(2)} \;,
\label{2.2}
\end{equation}

\noindent 
where $[T_{\pi N}^{(m)}]^{(i)}$ is the $\pi N$ amplitude for nucleon 
$(i)$ expanded at order $O(q^m)$. The factor within curly brackets 
in the integrand is $O(q^0)$ whereas the leading term in $T_{\pi N}$, 
as given by the Weinberg-Tomozawa theorem \cite{W66,T66}, is $O(q)$. 
Thus ${\cal T}_{T\!P}^{(n)}$ requires $T_{\pi N}$ up to $O(q^{n-1})$. 

This result is important regarding the numerical values of the LECs to be 
used in the determination of the TPEP, which depend on the chiral order 
one is working at \cite{MK}. These constants are not observables and must 
be obtained from empirical quantities such as, for instance, $\pi N$ 
subthreshold coefficients. In the case of our $O(q^4)$ TPEP, 
consistency demands 
the use of LECs determined from $T_{\pi N}$ at $O{(q^{3})}$. 

Finally, a further motivation for deriving the TPEP from the 
intermediate $\pi N$ amplitude is that this stresses the continuity of 
present developments with the seminal works of the Paris \cite{P} and 
Stony Brook \cite{SB} groups, produced more than three decades ago. For 
this very reason, one becomes better prepared to understand the specific 
role played by ChPT in this problem. 

\section{configuration space potential} \label{secIII}

The configuration space Schr\"odinger equation is a rather useful tool 
for calculating low energy nuclear processes. In principle, the 
$O(q^4)$ $r$-space potential could be obtained by just performing 
the Fourier transform of our center of mass  $p$-space potential, 
which is written as\footnote{In this result, the (+) and (-) 
upper labels indicate, respectively, terms arising from the isospin even 
and odd $\pi N$ subamplitudes.} 

\begin{equation}
{t}_{cm}=3\;{t}^++2\;{\mbox{\boldmath $\tau$}}^{(1)} 
\cdot{\mbox{\boldmath $\tau$}}^{(2)}\;{t}^-
\label{3.1}
\end{equation}  

\noindent
with 

\begin{equation}
{t}_{cm}^{\pm}
={t}_C^\pm
+\frac{{\mbox{\boldmath $\Omega$}}_{LS}}{m^2}\; {t}_{LS}^\pm
+\frac{{\mbox{\boldmath $\Omega$}}_{T}}{m^2}\; {t}_{T}^\pm
+\frac{{\mbox{\boldmath $\Omega$}}_{SS}}{m^2}\; {t}_{SS}^\pm
+\frac{{\mbox{\boldmath $\Omega$}}_{Q}}{m^4}\;{t}_{Q}^\pm\;,
\label{3.2}
\end{equation}

\noindent
and $\;{\mbox{\boldmath $\Omega$}}_{LS}=
i\;({\mbox{\boldmath $\sigma$}}^{(1)}\!+\!{\mbox{\boldmath $\sigma$}}^{(2)})
\cdot {\mbox{\boldmath $q$}}\times{\mbox{\boldmath $z$}}/4$, 
${\mbox{\boldmath $\Omega$}}_T=-{\mbox{\boldmath $q$}}^2\;
(3{\mbox{\boldmath $\sigma$}}^{(1)}\cdot{\mbox{\boldmath $\hat{q}$}}
{\mbox{\boldmath $\sigma$}}^{(2)}\cdot{\mbox{\boldmath $\hat{q}$}}\!
-\!{\mbox{\boldmath $\sigma$}}^{(1)}\cdot{\mbox{\boldmath $\sigma$}}^{(2)})$, 
${\mbox{\boldmath $\Omega$}}_{SS}={\mbox{\boldmath $q$}}^2\;
{\mbox{\boldmath $\sigma$}}^{(1)}\cdot{\mbox{\boldmath $\sigma$}}^{(2)}$, 
and 
${\mbox{\boldmath $\Omega$}}_Q={\mbox{\boldmath $\sigma$}}^{(1)}\cdot
{\mbox{\boldmath $q$}}\!\times\!{\mbox{\boldmath $z$}} \;
{\mbox{\boldmath $\sigma$}}^{(2)}\cdot{\mbox{\boldmath $q$}}\!
\times\!{\mbox{\boldmath $z$}}$. 
However, this leads to expressions that contain non-local terms, due 
to the energy dependence of the profile functions ${t}_i$. In order 
to avoid this kind of complication, we expand the potential in the 
non-local operators and keep only local and spin-orbit contributions. 
In this approximation, the configuration space potential becomes 

\begin{eqnarray}
V({\mbox{\boldmath $r$}})&=&
\left(V^{+}_{C}+V^{+}_{LS}\,\Omega_{LS}+V^{+}_{T}\,\Omega_{T}
+V^{+}_{SS}\,\Omega_{SS}\right)
\nonumber\\[2mm]
&+&{\mbox{\boldmath $\tau$}}^{(1)}\cdot{\mbox{\boldmath $\tau$}}^{(2)}
\left(V^{-}_{C}+V^{-}_{LS}\,\Omega_{LS}+V^{-}_{T}\,\Omega_{T}
+V^{-}_{SS}\,\Omega_{SS}\right)
\label{3.3}
\end{eqnarray}

\noindent
with $\; \Omega_{LS}= {\mbox{\boldmath $L$}} \cdot 
({\mbox{\boldmath $\sigma$}}^{(1)}+{\mbox{\boldmath $\sigma$}}^{(2)})/2 ,
\;\;\Omega_{T}= 3\, {\mbox{\boldmath $\sigma$}}^{(1)}\cdot 
\hat{{\mbox{\boldmath $r$}}} \; {\mbox{\boldmath $\sigma$}}^{(2)}\cdot 
\hat{{\mbox{\boldmath $r$}}} - {\mbox{\boldmath $\sigma$}}^{(1)}\cdot 
{\mbox{\boldmath $\sigma$}}^{(2)},\;\; \Omega_{SS} = 
{\mbox{\boldmath $\sigma$}}^{(1)}\cdot {\mbox{\boldmath $\sigma$}}^{(2)}.$

The radial functions are given by 

\begin{eqnarray}
V_{C}^{\pm}(r) &=& \tau^{\pm}\;U_{C}^{\pm}(x)\,,
\label{3.4}\\[2mm]
V_{LS}^{\pm}(r) &=& \tau^{\pm}\;\frac{\mu^2}{m^2}\,\frac{1}{x}\,\frac{d}{dx}\,
U_{LS}^{\pm}(x)\,,
\label{3.5}\\[2mm]
V_{T}^{\pm}(r) &=& \tau^{\pm}\;\frac{\mu^2}{m^2}\left[\frac{d^2}{d x^2}
-\frac{1}{x}\,\frac{d}{dx}\right]\,U_{T}^{\pm}(x)\,,
\label{3.6}\\[2mm]
V_{SS}^{\pm}(r) &=& -\tau^{\pm}\;\frac{\mu^2}{m^2}\left[\frac{d^2}{dx^2}
+\frac{2}{x}\,\frac{d}{dx}\right]\,U_{SS}^{\pm}(x)\,,
\label{3.7}
\end{eqnarray}

\noindent
where $\tau^+=3$, $\tau^-=2$, $x={\mu} r$, and 

\begin{equation}
U^\pm_I(x) = -\int \frac{d^3k}{(2\pi)^3}\, e^{i\,{\mbox{\boldmath $k$}} 
\cdot {\mbox{\boldmath $x$}} }\; {t}^\pm_I(k) \,,
\qquad\qquad I=\{C,LS,T,SS\}
\label{3.8}
\end{equation} 

\noindent with ${\mbox{\boldmath $k$}}={\mbox{\boldmath $q$}}/\mu$. This 
allows the potential to be expressed in terms of dimensionless 
configuration space Feynman integrals, denoted by $S$, and related to the 
functions $\Pi$ of Ref.\cite{HR} by 

\begin{equation}
S(x)=\int\frac{d^3k}{(2\pi)^3}\,e^{i\,{\mbox{\boldmath $k$}}\cdot
{\mbox{\boldmath $x$}}}\;\Pi({\mbox{\boldmath $k$}})\,.
\end{equation}

Using the results of Sec. IX of Ref.\cite{HR}, we have the 
expansions\footnote{In 
writing these expressions, we did not consider the relativistic 
normalization factor, proportional to $m/E$.} 

\begin{eqnarray}
&& U_C^+ =-\frac{\mu^3m^2}{256\pi^2f_{\pi}^4}\;\left[
\frac{{\mu}}{m}\right]^2 \left\{ g_A^4\,(1+4\,\Delta_{GT})\;(1\!
-\!{\mbox{\boldmath $\nabla$}}^2/2)^2 \left( S_\times-S_b\right)\right.
\nonumber\\[2mm]
&& \left. + \left[ \frac{{\mu}}{m}\right] g_A^2\;(1\!
-\!{\mbox{\boldmath $\nabla$}}^2/2) \left[ - g_A^2 
\left( 2 S_a + {\mbox{\boldmath $\nabla$}}^2 S_t \right)
+ 8\left( \bar{\delta}_{00}^+ + \bar{\delta}_{01}^+ 
{\mbox{\boldmath $\nabla$}}^2 \right) S_t \right]
\right.
\nonumber\\[2mm]
&& \left. + \left[ \frac{{\mu}}{m}\right]^2 \left[
- \frac{m^2\,g_A^4}{16\pi^2f_{\pi}^2}\; (1\!-\!2 
{\mbox{\boldmath $\nabla$}}^2)\left( - 4\pi (1\!
-\!{\mbox{\boldmath $\nabla$}}^2/2) S_t + (1\!
-\!{\mbox{\boldmath $\nabla$}}^2/2)^2 S_{tt} \right) 
+\frac{g_A^4}{4}{\mbox{\boldmath $\nabla$}}^2 (S_\times+S_b)\right]
\right.
\nonumber\\[2mm]
&& \left. 
+ \left[ \frac{{\mu}}{m}\right]^2 \left[ 
g_A^4 {\mbox{\boldmath $\nabla$}}^4-4 g_A^2  \left(( \bar{\delta}_{00}^+ 
+ \bar{\delta}_{01}^+ {\mbox{\boldmath $\nabla$}}^2) 
{\mbox{\boldmath $\nabla$}}^2+ \delta_{10}^+ 
(1-2{\mbox{\boldmath $\nabla$}}^2/3 + {\mbox{\boldmath $\nabla$}}^4/6) \right)
\right.\right.\nonumber\\[2mm]&&\left.\left. 
+8\left( \bar{\delta}_{00}^+ +\bar{\delta}_{01}^+ 
{\mbox{\boldmath $\nabla$}}^2 + ({\delta}_{10}^+/3)
(1 - {\mbox{\boldmath $\nabla$}}^2/4)\right)^2
+\frac{32}{45}({\delta}_{10}^+)^2 (1\!
-\! {\mbox{\boldmath $\nabla$}}^2)^2\right] S_\ell\right\}\;,
\label{3.9}\\[5mm]
&& U_{LS}^+ = -\frac{\mu^3m^2g_A^2}{128\pi^2f_{\pi}^4}\;
\left[ \frac{{\mu}}{m }\right] \left\{ g_A^2 
\left[(1- {\mbox{\boldmath $\nabla$}}^2/2)(\tilde S_b-S_t)
-(3/2-5{\mbox{\boldmath $\nabla$}}^2/8)\,S_a\right] 
\right.\nonumber\\[2mm]&&\left.
+ \left[\frac{{\mu}}{m}\right] \left[ \frac{g_A^2}{4}\,
(1+2{\mbox{\boldmath $\nabla$}}^2 - {\mbox{\boldmath $\nabla$}}^4/2)\; 
(S_\times+S_b)+\left( 2\,g_A^2 {\mbox{\boldmath $\nabla$}}^2 
- \frac{16}{3}\,\delta^+_{10}\,(1\!-\!{\mbox{\boldmath $\nabla$}}^2/4)\right) 
S_\ell \right] \right\}\;,
\label{3.10}\\[5mm]
&&  U_T^+ =  \frac{U_{SS}^+}{2} =-
\frac{\mu^3m^2g_A^2}{768\pi^2f_{\pi}^4}\; \left\{ - g_A^2\,(1+4\,\Delta{GT})\; 
(1\!-\!{\mbox{\boldmath $\nabla$}}^2/4)\left[ S_\times + S_b\right] 
\right.\nonumber\\[2mm]&&\left.
- \left[\frac{{\mu}}{m}\right] \frac{g_A^2}{2}\left[ (1
- {\mbox{\boldmath $\nabla$}}^2/2)(S_t-\tilde S_b)
+(1- {\mbox{\boldmath $\nabla$}}^2/4)S_a \right]
\right.\nonumber\\[2mm]&&\left.
+\left[\frac{{\mu}}{m} \right]^2  \left[ \frac{g_A^2}{4}\;(1\!
-\!{\mbox{\boldmath $\nabla$}}^2/2)^2\; S_\times 
+ \frac{4}{3}\,\beta^+_{00}(1 - {\mbox{\boldmath $\nabla$}}^2/4) 
S_\ell \right]\right\} \;,
\label{3.11}
\end{eqnarray}

\noindent
and

\begin{eqnarray}
&&  U_C^- =-\frac{\mu^3m^2}{16\pi^2f_{\pi}^4} 
\left[ \frac{{\mu}}{m }\right]^2\left\{
\frac{g_A^4}{16}\,(1+4\,\Delta_{GT})\,
(1-{\mbox{\boldmath $\nabla$}}^2/2)^2 (S_\times+S_b) 
\right.\nonumber\\[2mm]&&\left.
-\frac{1}{4}\left[g_A^4\,(1+4\,\Delta_{GT})-g_A^2\,(1+2\,\Delta_{GT})\right]
(1-{\mbox{\boldmath $\nabla$}}^2/2)S_\ell
\right.\nonumber\\[2mm]&&\left.
+\frac{1}{24}
\left[g_A^4\,(1+4\,\Delta_{GT})-2g_A^2\,(1+2\,\Delta_{GT})+1\right]
(1-{\mbox{\boldmath $\nabla$}}^2/4)S_\ell
\right.\nonumber\\[2mm]&&\left.
+ \left[ \frac{{\mu}}{m}\right] \frac{g_A^2}{8} (1 
- {\mbox{\boldmath $\nabla$}}^2/2)\left[ g_A^2 (S_a 
- {\mbox{\boldmath $\nabla$}}^2 S_t /2) + (g_A^2-1) 
(1 - {\mbox{\boldmath $\nabla$}}^2/2) S_t \right]
\right.\nonumber\\[2mm]&&\left.
+ \left[ \frac{{\mu}}{m} \right]^2 
\left\{ \frac{g_A^2}{2} (1 - {\mbox{\boldmath $\nabla$}}^2/2) 
\left[  - (g_A^2-1) {\mbox{\boldmath $\nabla$}}^2/8 +\bar\delta^-_{00} 
+ \bar\delta^-_{01} {\mbox{\boldmath $\nabla$}}^2 + \delta^-_{10} 
(1 - {\mbox{\boldmath $\nabla$}}^2/4)/3
+ \bar\beta^-_{00}{\mbox{\boldmath $\nabla$}}^2/4 \right] 
\right.\right.\nonumber\\[2mm]&&\left.\left.
- \frac{(g_A^2-1)}{6} (1 - {\mbox{\boldmath $\nabla$}}^2/4) 
\left[ \bar\delta^-_{00} + \bar\delta^-_{01}{\mbox{\boldmath $\nabla$}}^2
+ 3\,\delta^-_{10} (1 - {\mbox{\boldmath $\nabla$}}^2/4)/5 
+ \bar\beta^-_{00}{\mbox{\boldmath $\nabla$}}^2/4 \right] \right\} S_\ell
\right.\nonumber\\[2mm]&&\left.
- \left[ \frac{{\mu}}{m} \right]^2 \frac{m^2}{64\pi^2f_{\pi}^2}
\left[ \left( 2 g_A^4  (1 - 5{\mbox{\boldmath $\nabla$}}^2/6 
+{\mbox{\boldmath $\nabla^4$}}/5) +4 (g_A^2-1)^2 
(1-3{\mbox{\boldmath $\nabla$}}^2/8+{\mbox{\boldmath $\nabla^4$}}/32)/9
\right.\right.\right.\nonumber\\[2mm] && \left.\left.\left.
- 4 g_A^2(g_A^2-1) (1-29{\mbox{\boldmath $\nabla$}}^2/72 
+ 7{\mbox{\boldmath $\nabla^4$}}/144) \right) S_\ell
\right.\right.\nonumber\\[2mm] && \left.\left.
\left( g_A^4 (1-{\mbox{\boldmath $\nabla$}}^2/2)^2 
- 2g_A^2(g_A^2-1)(1-{\mbox{\boldmath $\nabla$}}^2/2)
(1-{\mbox{\boldmath $\nabla$}}^2/4)/3 +(g_A^2-1)^2 
(1-{\mbox{\boldmath $\nabla$}}^2/4)^2 \right) S_{\ell \ell} \right]
\right.\nonumber\\[2mm]&&\left.
+ \left[ \frac{{\mu}}{m} \right]^2 \frac{g_A^4}{16} (1 
- {\mbox{\boldmath $\nabla$}}^2/2)^2 ({\mbox{\boldmath $\nabla$}}^2/4) 
(S_\times-S_b)\right\}\;,
\label{3.12}\\[5mm]
&& U_{LS}^- =-\frac{\mu^3m^2}{128\pi^2f_\pi^4}
\left[ \frac{{\mu}}{m}\right] \left\{ g_A^4 \left[ (3/2 
- 5 {\mbox{\boldmath $\nabla$}}^2/8)  S_a - (1 
- {\mbox{\boldmath $\nabla$}}^2/2) (S_t + \tilde S_b) \right]
\right.\nonumber\\[2mm]&&\left.
+ 2 g_A^2 (g_A^2-1) (1 - {\mbox{\boldmath $\nabla$}}^2/4) S_t
\right.\nonumber\\[2mm]&&\left.
+ \left[ \frac{{\mu}}{m}\right]  \left[ (g_A^2-1)^2 (1 
- {\mbox{\boldmath $\nabla$}}^2/4)/2 + 4\,g_A^2\;\bar\beta^-_{00}\;
(1 - {\mbox{\boldmath $\nabla$}}^2/2)- 4 \;(g_A^2-1) \bar\beta^-_{00}
(1 - {\mbox{\boldmath $\nabla$}}^2/4) /3 \right] S_\ell
\right.\nonumber\\[2mm]&&\left.
+ \left[ \frac{{\mu}}{m}\right] \frac{g_A^4}{4} (1 
- {\mbox{\boldmath $\nabla$}}^2/2)^2 (S_\times-S_b)
\right.\nonumber\\[2mm]&&\left.
- \left[ \frac{{\mu}}{m} \right] \frac{m^2 g_A^4}{8\pi^2f_\pi^2} 
\left[ -2\pi (1-{\mbox{\boldmath $\nabla$}}^2/4) S_t + (1 
- {\mbox{\boldmath $\nabla$}}^2/4)^2 S_{tt} \right]  \right\}\;,
\label{3.13}\\[5mm]
&& U_T^- =  \frac{U_{SS}^-}{2} =-\frac{\mu^3m^2}{1536\pi^2f_\pi^4}\;
\left[ \frac{{\mu}}{m }\right] \left\{ g_A^4 \left[ (1 
- {\mbox{\boldmath $\nabla$}}^2/2)\,\tilde S_b  + (1 
- {\mbox{\boldmath $\nabla$}}^2/4)\, S_a \right]
\right.\nonumber\\[2mm]&&\left.
- 2\,g_A^2 (g_A^2-1-2\,\bar\beta^-_{00}) 
(1 - {\mbox{\boldmath $\nabla$}}^2/4)\,S_t
\right.\nonumber\\[2mm]&&\left.
+ \left[\frac{{\mu}}{m}\right] \left[ -g_A^2 (g_A^2-1-2\,\bar\beta^-_{00}) 
(1 - {\mbox{\boldmath $\nabla$}}^2/2)- (g_A^2-1-2\,\bar\beta^-_{00})^2 
(1 - {\mbox{\boldmath $\nabla$}}^2/4) /3 \right] S_\ell
\right.\nonumber\\[2mm]&&\left.
+\left[\frac{{\mu}}{m}\right]\frac{m^2g_A^4}{8\pi^2f_\pi^2}
\left[-2\pi (1-{\mbox{\boldmath $\nabla$}}^2/4)S_t+(1
-{\mbox{\boldmath $\nabla$}}^2/4)^2 S_{tt}\right]\right\}\;,
\label{3.14}
\end{eqnarray}

\noindent where the Laplacians act on the variable $x$. 
The chiral orders of the various radial functions may be read directly 
from the combination of Eqs. (\ref{3.4})--(\ref{3.7}) 
and (\ref{3.9})--(\ref{3.14}). Their relative importances will be 
discussed in detail in Sec. \ref{secV}. 
We have expressed our results in terms of the axial coupling $g_A$. 
If one wants, they may be rewritten using the $\pi N$ coupling constant 
$g$, by means of the relation $g=(1+\Delta_{GT})\,g_A\,m/f_{\pi}$, 
where $\Delta_{GT}$ is the so called Goldberger-Treiman discrepancy. 

The parameters $\delta_{ij}^\pm$ and $\beta_{ij}^\pm$ entering these 
expressions are determined by subthreshold $\pi N$ coefficients or, 
alternatively, by the LECs of the effective Lagrangian, according to the 
results presented in Sec. V of Ref.\cite{HR}. Their empirical values 
are reproduced in Table \ref{tab1}. 

\begin{table}[b]
\begin{center}
\caption{Dimensionless subthreshold coefficients; definitions are 
the same as in Ref.[\ref{reftab1}].\label{tab1}}
\begin{tabular} {|c|c|c||c|}
\hline
\raisebox{0pt}[12pt][6pt]{}
$\bar{\delta}_{00}^+$&$\delta_{10}^+$&$\bar{\delta}_{01}^+$ & $\beta_{00}^+$
\\\hline
\raisebox{0pt}[12pt][6pt]{} -4.72 & 3.34 & 4.15 & -10.57 \\
\hline\hline
\raisebox{0pt}[12pt][6pt]{}
$\bar{\delta}_{00}^-$&$\delta_{10}^-$&$\bar{\delta}_{01}^-$&$\bar\beta_{00}^-$
\\ \hline
\raisebox{0pt}[12pt][6pt]{} 7.02 & -3.35 &-2.05 & 5.04 \\ \hline
\end{tabular}
\end{center}
\end{table}

The eight functions $S_i$ which carry the spatial dependence 
of the potential are dimensionless and given by 

\begin{eqnarray}
S_\ell &=& \frac{K_1(2x)}{\pi x^2}\;,
\label{3.15}\\[2mm]
S_a &=& -\frac{e^{-2x}}{2x^2}\;,
\label{3.16}\\[2mm]
S_t&=&-\int_0^1da\int_0^1 db\;
\frac{(1-b)\;2m/\mu}{\Lambda_t^2}\;\frac{e^{-\theta_t\,x}}{4\pi x}\;,
\label{3.17}\\[2mm]
&&\Lambda_t^2=a(1\!-\!a)(1\!-\!b)^2\;,
\nonumber\\
&&\theta_t^2=\left[(1-b)+b^2m^2/{\mu}^2\right]/\Lambda_t^2\;,
\nonumber\\
S_\times &=& \int_0^1 \!db \int_0^1 \!d a \; 
\frac{a^2b\;\;\; 4m^2/{\mu}^2}{\Lambda_\times^4}\;
\frac{e^{-{\theta}_\times x}}{8\pi {\theta}_\times} \;,
\label{3.20}\\[2mm]
&&\Lambda_\times^2=a(1\!-\!a)(1\!-\!b)\;,
\nonumber\\
&&\theta_\times^2=\left[(1-ab)+a^2b^2m^2/{\mu}^2\right]/\Lambda_\times^2\;,
\nonumber\\
S_b &=& \int_1^\infty \!dc \int_0^1 \!db \int_0^1 \!d a \; 
\frac{a^2b\;\;\; 4m^2/{\mu}^2}{\Lambda_b^4}\;
\frac{e^{-{\theta}_b x}}{8\pi {\theta}_b} \;,
\label{3.21}\\[2mm]
&&\Lambda_b^2 = a(1-a)(1-b) - a^2 b^2 (1-c^2) /4 \;,
\nonumber\\
&&\theta_b^2 = \left[ (1-ab) + a^2 b^2 c^2 (m/{\mu})^2 \right] / \Lambda_b^2 \;,
\nonumber\\
{\tilde{S}}_b  &=& \int_1^\infty \!dc \int_0^1 \!db \int_0^1 \!d a \; 
\frac{a^3b^2\;\;\; 4m^3/{\mu}^3}{\Lambda_b^4}\;
\frac{e^{-{\theta}_b  x}}{8\pi {\theta}_b} \;,
\label{3.24}\\[2mm]
S_{\ell\ell} &=& 
-\frac{1}{4\pi}\int_0^1da\int_0^1db\,\frac{\sqrt{(1\!-\!a)(1\!-\!b)}}{(b+a)}\,
\frac{1}{x}\,\frac{1}{\left(\frac{1}{a}\!-\!\frac{1}{b}\right)}\left[
\left(\frac{4}{a^2}\right)^2\frac{e^{-2x/a}}{x}
\!-\!\left(\frac{4}{b^2}\right)^2\frac{e^{-2x/b}}{x}\right]\;,
\label{3.25}\\[2mm]
S_{tt} &=& -\frac{(4m\mu)^2}{\pi}\int_0^1da\int_0^1db\,
\frac{G(4\mu^2/a)\,G(4\mu^2/b)}{a^2b^2(a+b)}\,\frac{1}{x}\,
\frac{1}{\left(\frac{1}{a}\!-\!\frac{1}{b}\right)}
\left[e^{-2x/a}\!-\!e^{-2x/b}\right]\;,
\label{3.26}
\end{eqnarray}

\noindent where $K_1(x)$ is the modified Bessel function and 

\begin{equation}
G(t')=\frac{2}{m\sqrt{t'(4m^2-t')}}\,\arctan
\frac{\sqrt{(4m^2-t')(t'-4\mu^2)}}{t'-2\mu^2}\,.
\label{eq:Gdef}
\end{equation}

In Figs. \ref{fig:ch6-potVc+}--\ref{fig:ch6-potVss-} we display the 
numerical predictions of our TPEP (full line), obtained by using the 
parameters $\delta_{ij}^{\pm}$ and $\beta_{ij}^{\pm}$ given in 
table \ref{tab1}, fixed by 
the $\pi N$ subthreshold coefficients of Ref.\cite{H}. As we will 
discuss in the sequence, our chiral TPEP is theoretically reliable 
for large distances and definitely not valid for internucleon separations 
smaller than 1 fm (shaded area). For the sake of producing a feeling for 
the phenomenological implications of these results, we also plot the 
medium range components of the Av14 \cite{av14} and Av18 \cite{av18} 
versions of the Argonne potential (dotted and dashed lines respectively). 

The central isoscalar component of the nuclear force is by far the most 
important one and the fact that the chiral prediction is consistent with 
both Argonne versions is rather reassuring.\footnote{The leading structure 
of $V_C^+$ was discussed in Ref.\cite{R01}.} The assessment of the other 
components is more difficult, since there are important variations 
between the Av14 and Av18 results. In the cases of $V_{SS}^{+}$, 
$V_{T}^{-}$, $V_{SS}^{-}$, where these variations do not involve signs, 
it is possible to note a qualitative agreement with the behavior of the 
chiral TPEP. The curves for $V_{SS}^{+}$, $V_{C}^{-}$ and $V_{T}^{-}$ 
are not far from those of Av18 whereas $V_{SS}^{-}$ coincides with the 
Av14 prediction. 

In order to complete the long-distance description of the 
$NN$ potential, one has to include the OPEP, which contributes only 
to $V_{T}^{-}$ and $V_{SS}^{-}$, through the following 
expressions: 

\begin{eqnarray}
\left.V_{T}^{-}\right|_{OPEP}&=&-\frac{m}{E}\,\frac{\mu^3g_A^2}{48\pi m^2}
\,(1+2\,\Delta_{GT})\left(x^2+3x+3\right)\frac{e^{-x}}{x^3}\,,
\\[3mm]
\left.V_{SS}^{-}\right|_{OPEP}&=&\frac{m}{E}\,\frac{\mu^3g_A^2}{48\pi m^2}
\,(1+2\,\Delta_{GT})\,\frac{e^{-x}}{x}\,.
\end{eqnarray}

These components, which dominate at large distances, are shown in 
Figs. \ref{fig:ch6-potopepVt-} and \ref{fig:ch6-potopepVss-}, together 
with the corresponding TPEP contributions. The influence of the TPEP 
only becomes significant in $V_{T}^{-}$ for $r<$ 2 fm, and in $V_{SS}^{-}$, 
for $r<$ 3 fm.

\newcounter{renato}
\setcounter{renato}{\thefigure}
\addtocounter{renato}{1}
\renewcommand{\thefigure}{\arabic{renato}(\alph{figure})}
\setcounter{figure}{0}

%
\newpage
\begin{figure}[!tbp]
\caption{Central isoescalar component.}
\begin{center}
\epsfig{figure=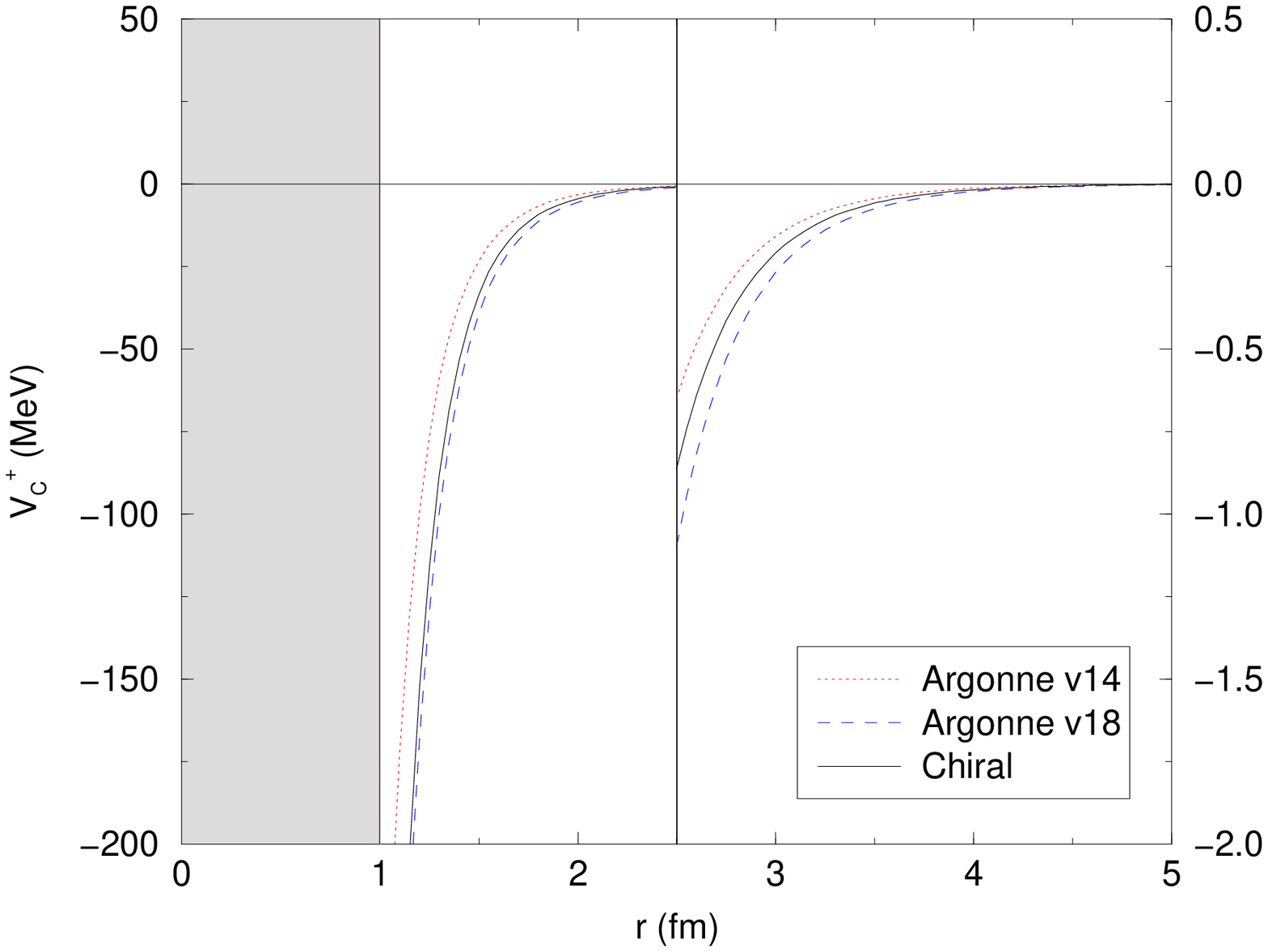, height=3.75in}
\end{center}
\label{fig:ch6-potVc+}
\end{figure}
\nopagebreak
\begin{figure}[!tbp]
\caption{Spin-orbit isoescalar component.}
\begin{center}
\epsfig{figure=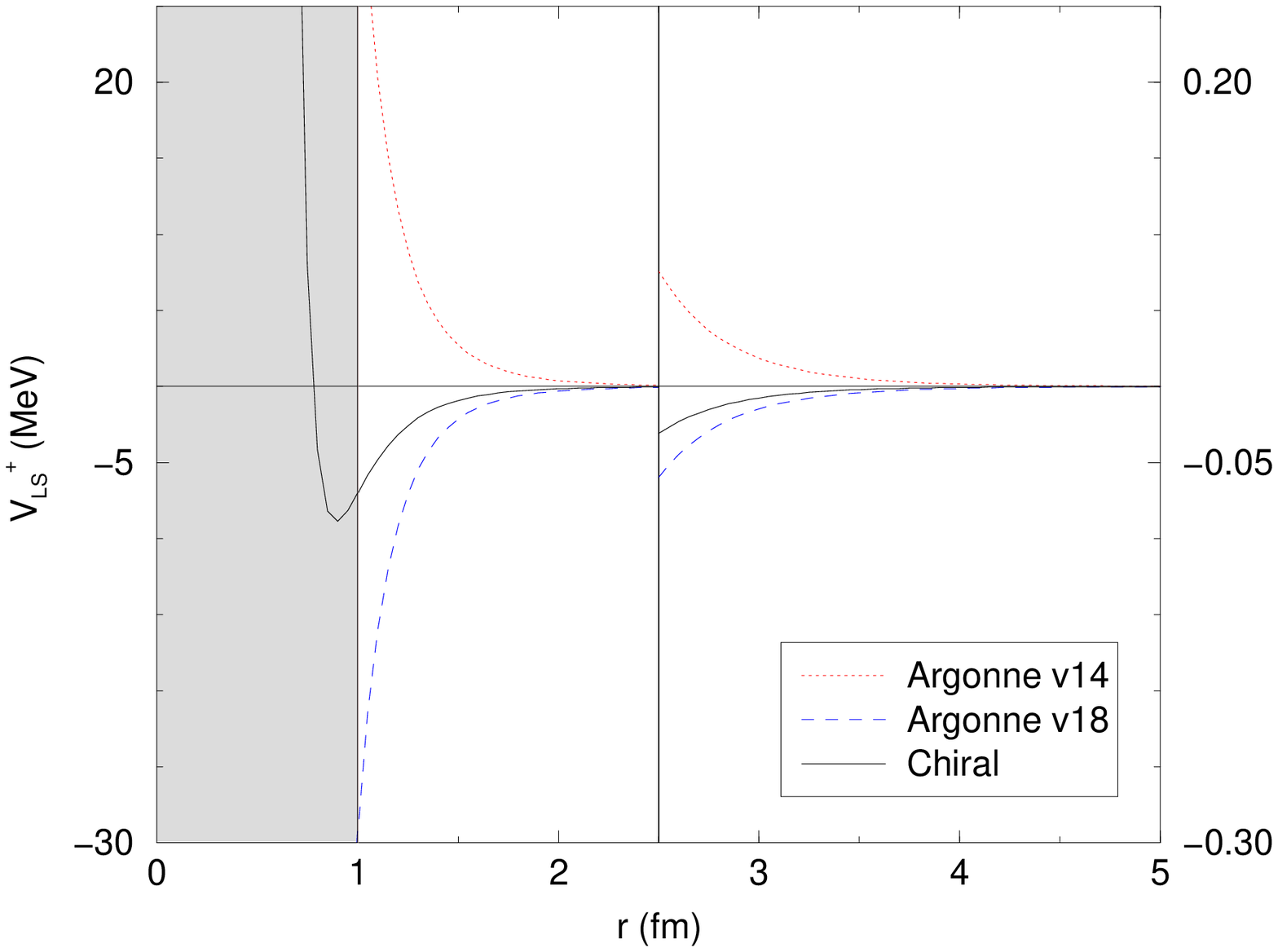, height=3.75in}
\end{center}
\label{fig:ch6-potVls+}
\end{figure}
\newpage
\begin{figure}[!tbp]
\caption{Tensor isoescalar component.}
\begin{center}
\epsfig{figure=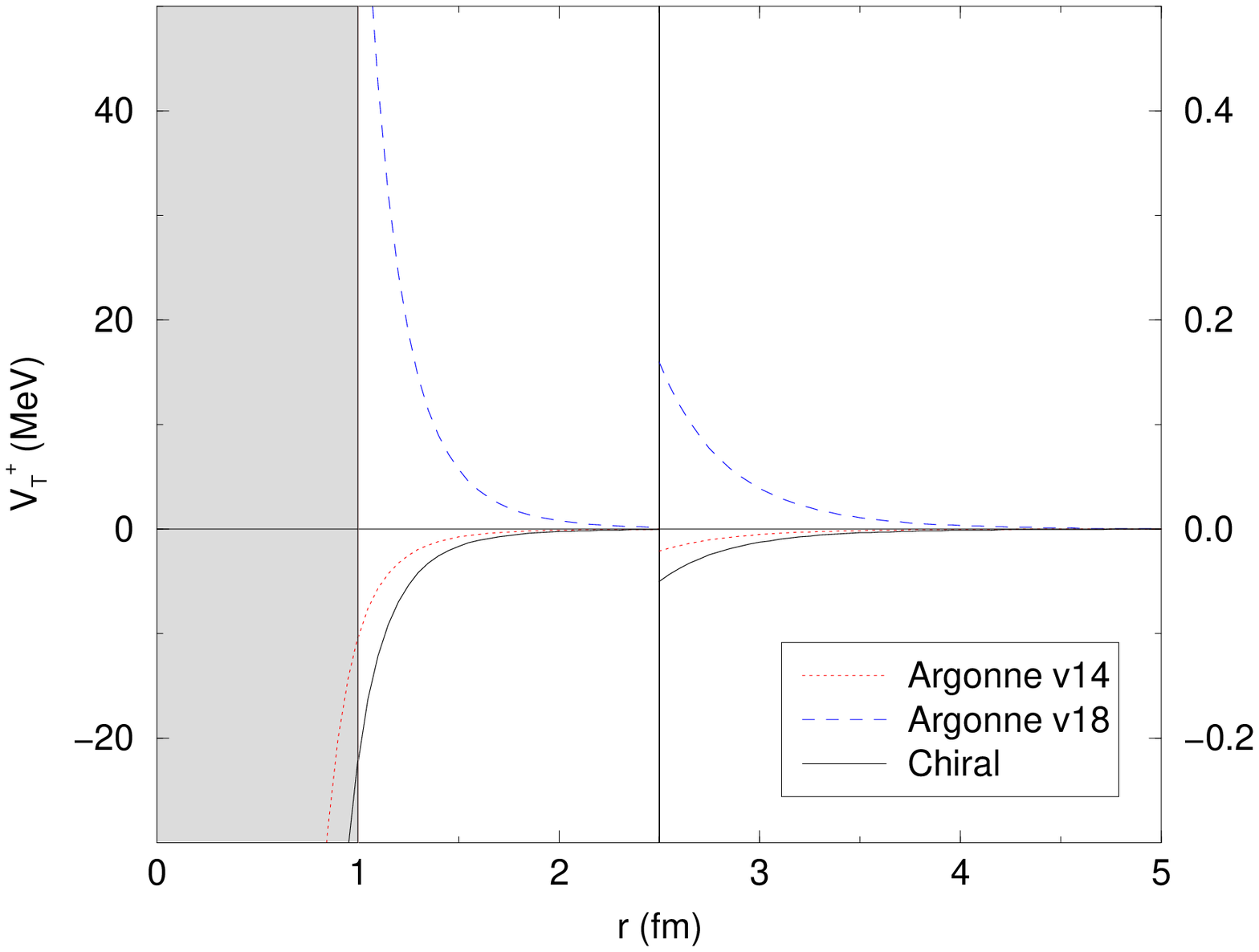, height=3.75in}
\end{center}
\label{fig:ch6-potVt+}
\end{figure}
\nopagebreak
\begin{figure}[!tbp]
\caption{Spin-spin isoescalar component.}
\begin{center}
\epsfig{figure=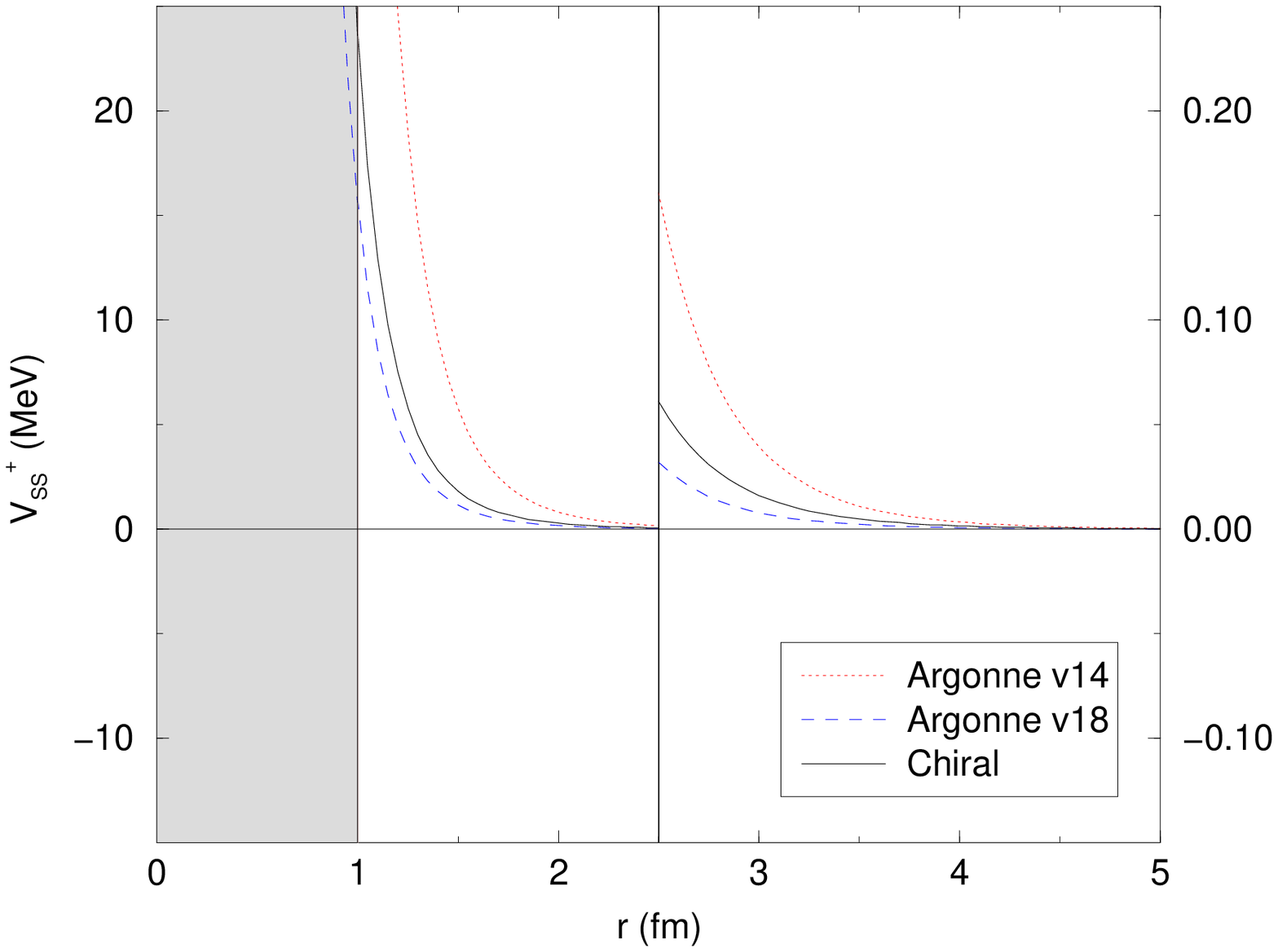, height=3.75in}
\end{center}
\label{fig:ch6-potVss+}
\end{figure}
\newpage
\begin{figure}[!tbp]
\caption{Central isovector component.}
\begin{center}
\epsfig{figure=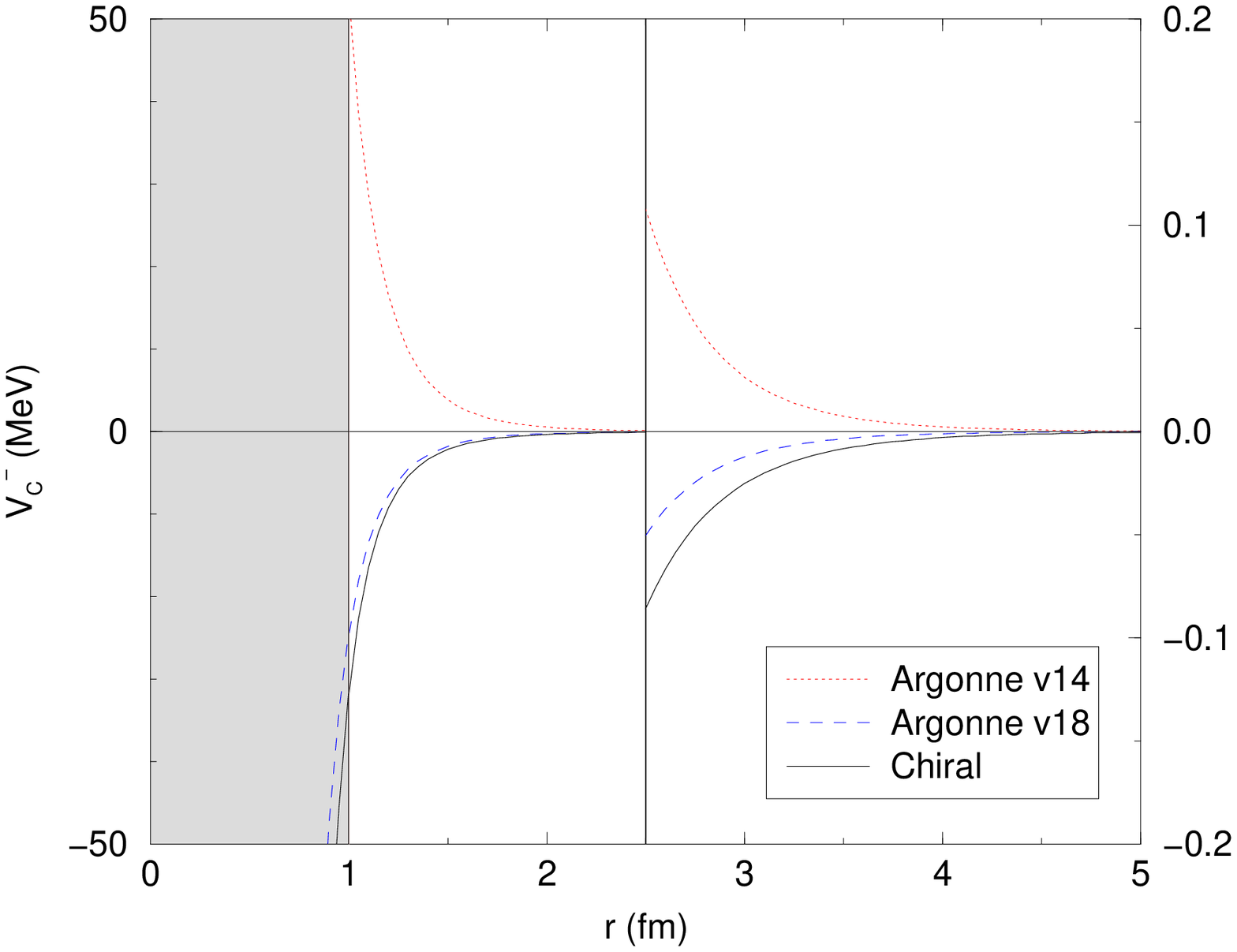, height=3.75in}
\end{center}
\label{fig:ch6-potVc-}
\end{figure}
\nopagebreak
\begin{figure}[!tbp]
\caption{Spin-orbit isovector component.}
\begin{center}
\epsfig{figure=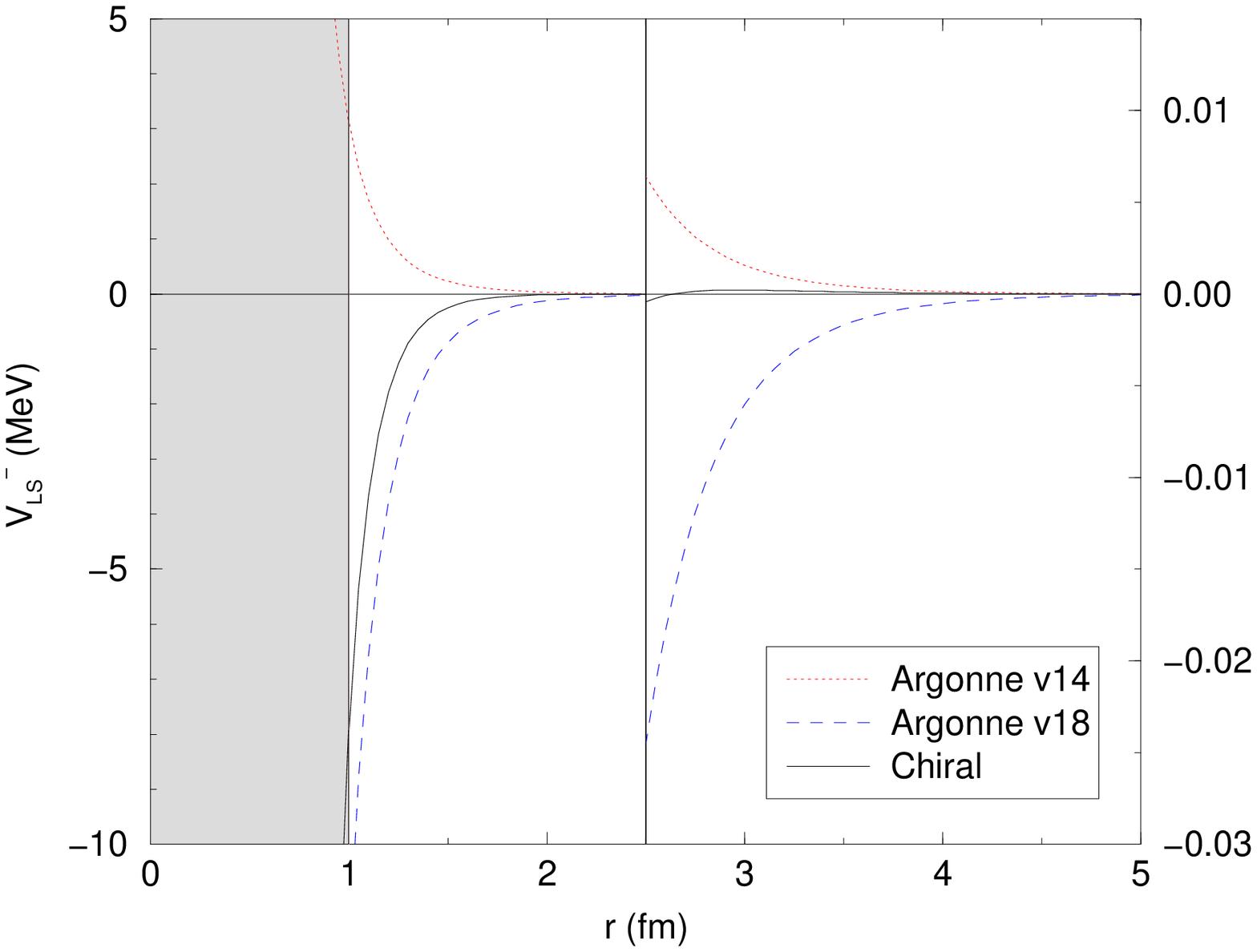, height=3.75in}
\end{center}
\label{fig:ch6-potVls-}
\end{figure}
\newpage
\begin{figure}[!tbp]
\caption{Tensor isovector component.}
\begin{center}
\epsfig{figure=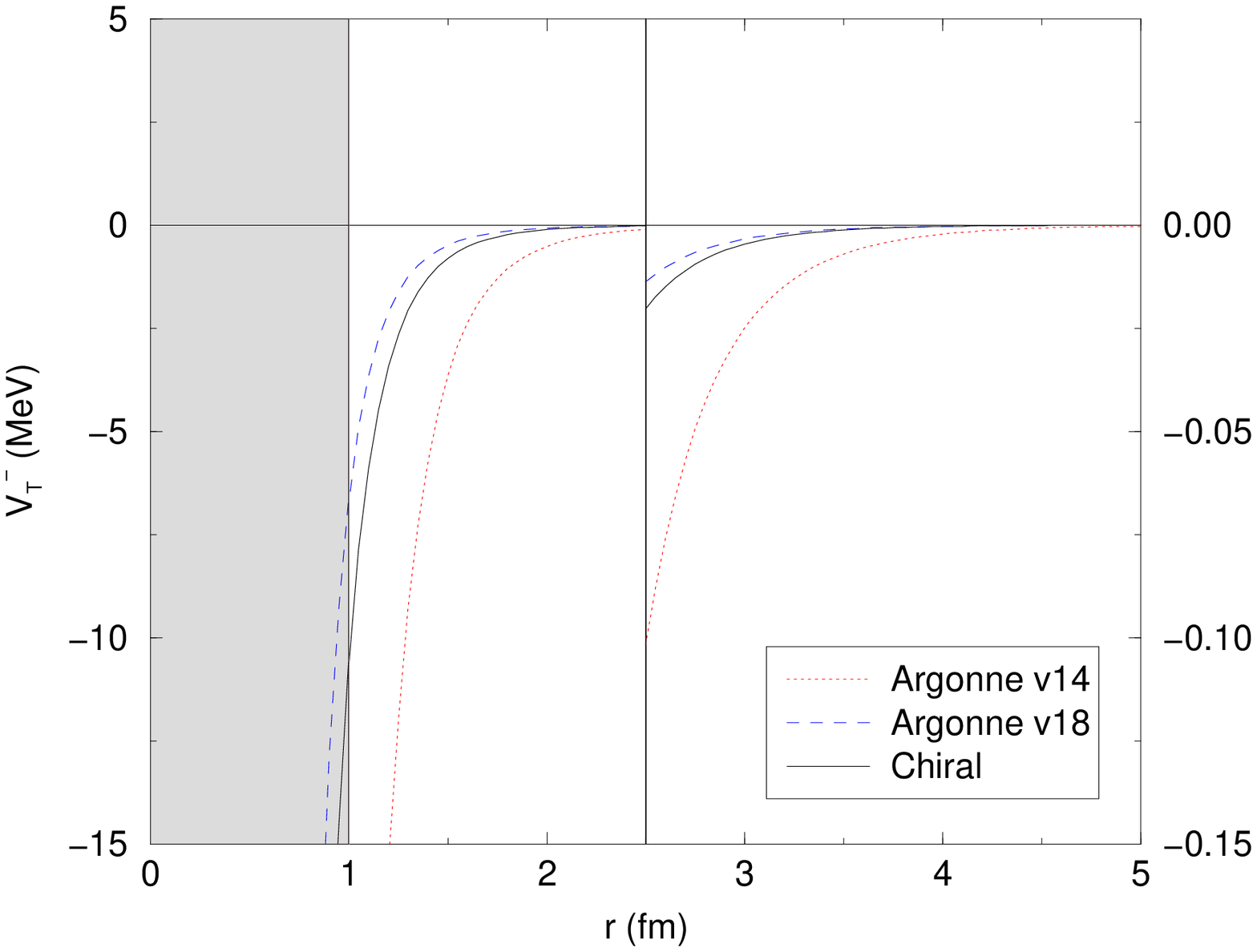, height=3.75in}
\end{center}
\label{fig:ch6-potVt-}
\end{figure}
\nopagebreak
\begin{figure}[!tbp]
\caption{Spin-spin isovector component.}
\begin{center}
\epsfig{figure=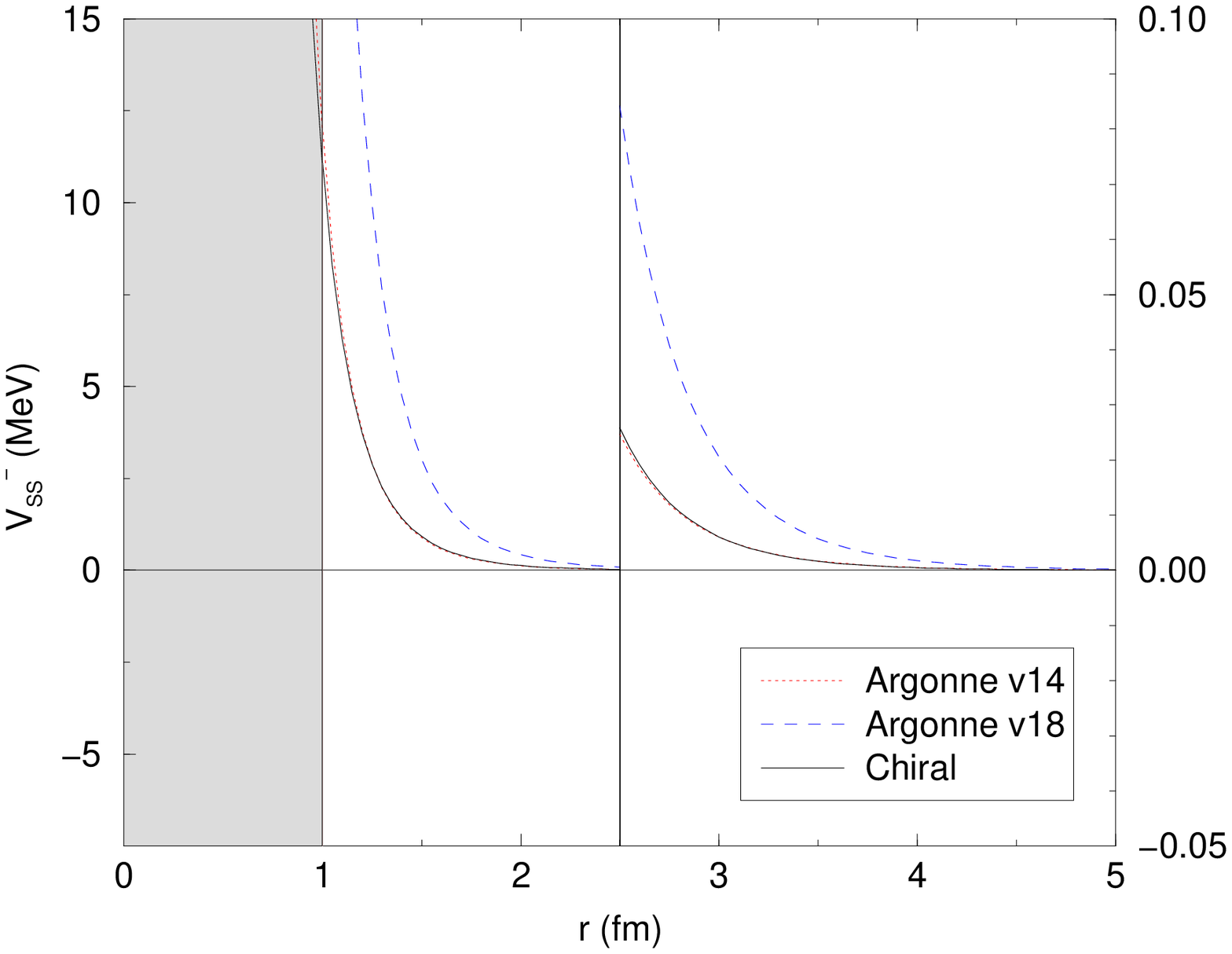, height=3.75in}
\end{center}
\label{fig:ch6-potVss-}
\end{figure}

\setcounter{figure}{\therenato}
\renewcommand{\thefigure}{\arabic{figure}}


\setcounter{renato}{\thefigure}
\addtocounter{renato}{1}
\renewcommand{\thefigure}{\arabic{renato}.(\alph{figure})}
\setcounter{figure}{0}

\newpage
\begin{figure}[!tbp]
\caption{OPEP and TPEP contributions to the tensor isovector potential.}
\begin{center}
\epsfig{figure=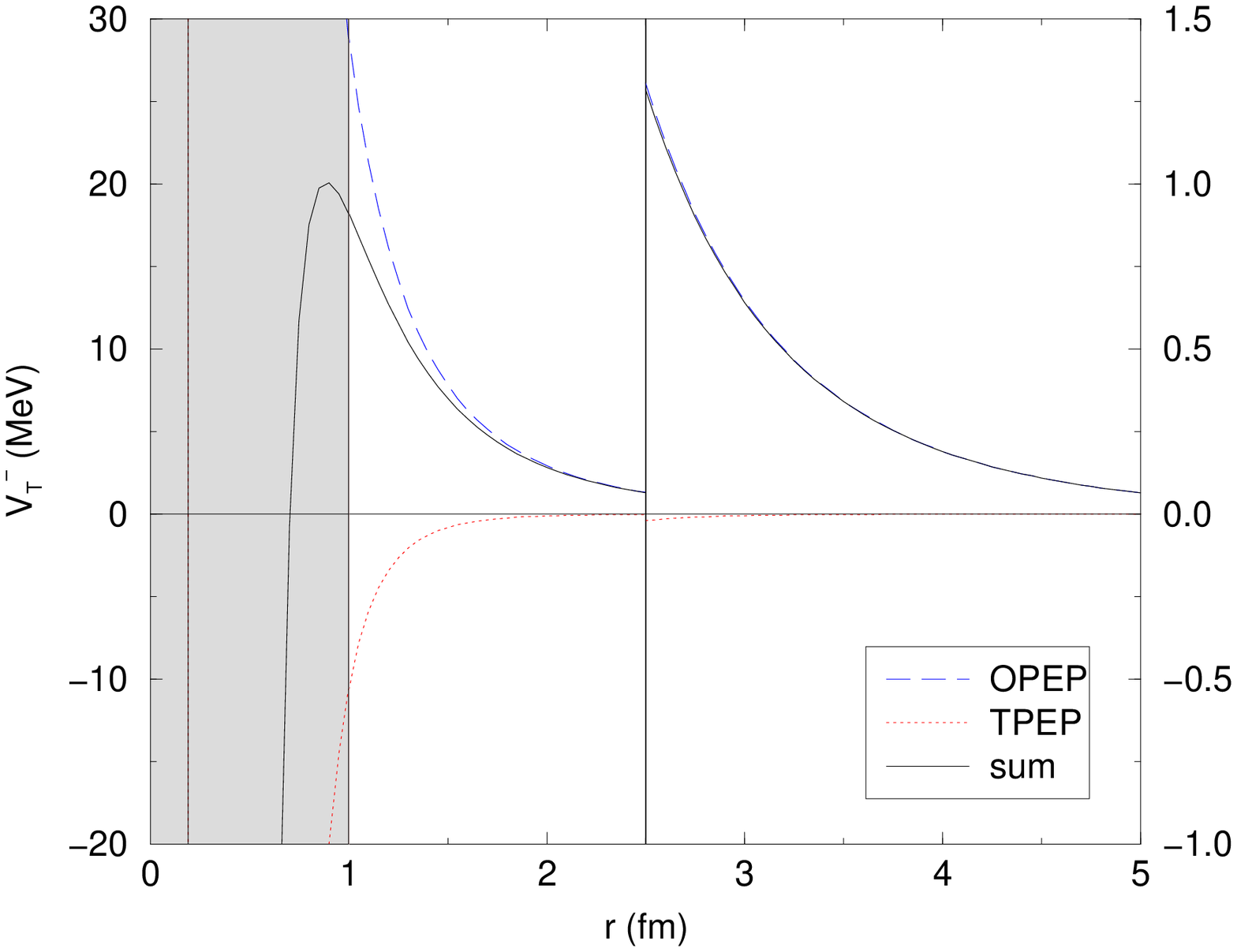, height=3.75in}
\end{center}
\label{fig:ch6-potopepVt-}
\end{figure}
\nopagebreak
\begin{figure}[!tbp]
\caption{OPEP and TPEP contributions to the spin-spin isovector potential.}
\begin{center}
\epsfig{figure=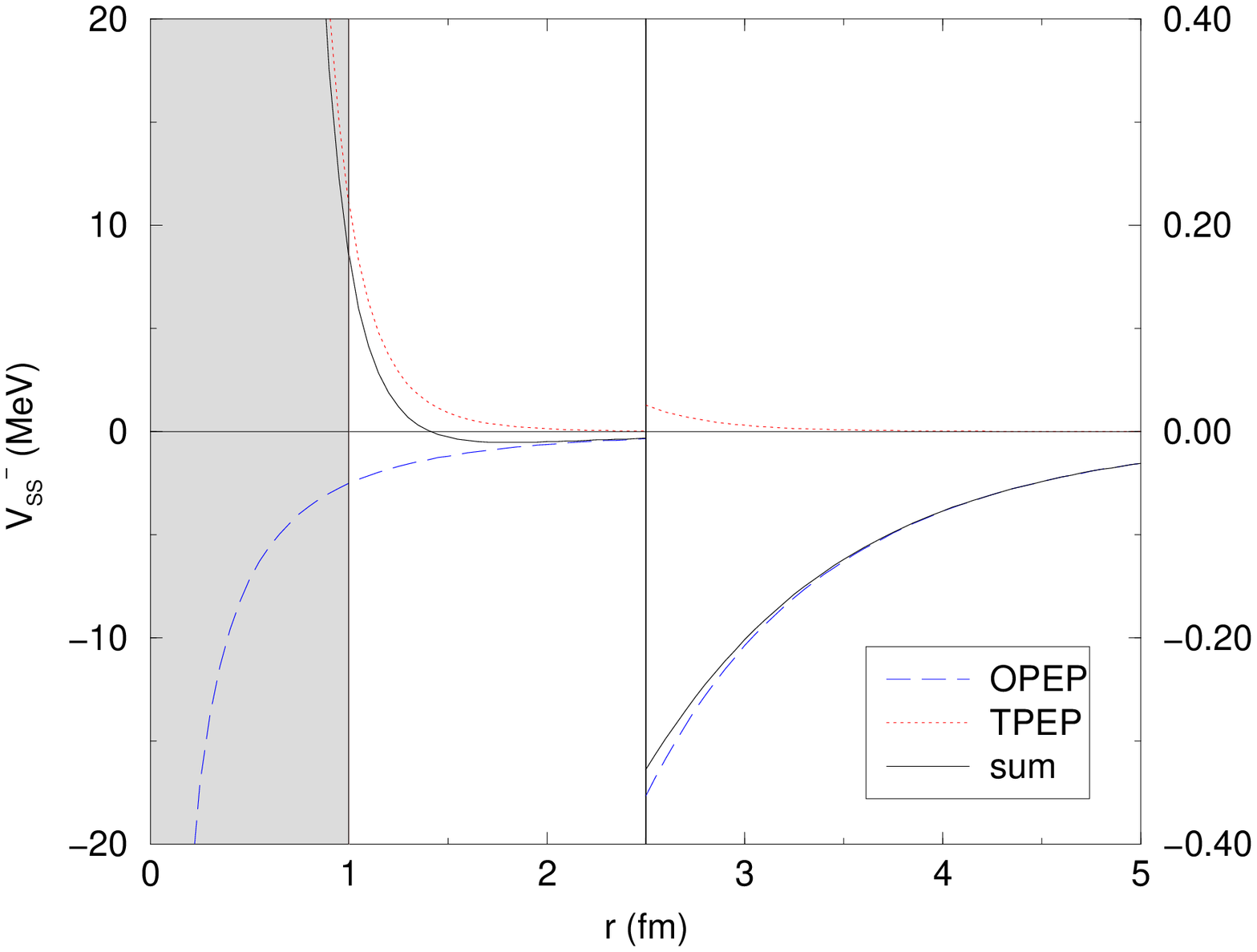, height=3.75in}
\end{center}
\label{fig:ch6-potopepVss-}
\end{figure}

\setcounter{figure}{\therenato}
\renewcommand{\thefigure}{\arabic{figure}}

\section{internal dynamics} \label{secIV}

In this section we discuss the relative importance of the contributions 
originating from the three families of diagrams presented in Fig. \ref{fig2}. 
This is motivated by the fact that the chiral description of the TPEP 
consists of a well defined field theoretical structure which depends on 
external parameters representing masses ($\mu$, $m$), coupling constants 
($f_{\pi}$, $g_A$), and LECs ($c_i$, $d_i$). In order to be able to 
obtain predictions, one has to feed the mathematical structure with the 
empirical values of these parameters. 

The constants present in the $O(q^4)$ potential 
may be divided into two classes, according to their numerical accuracy. 
The values of $\mu$, $m$, $f_{\pi}$, and $g_A$ entering 
${\cal L}_{\pi}^{(2)}$ and ${\cal L}_{N}^{(1)}$ may be considered as being 
very precise for the purposes of determining the TPEP. On the other hand, 
the constants $c_i$ and $d_i$ that appear in ${\cal L}_{N}^{(2)}$ and 
${\cal L}_{N}^{(3)}$ need to be extracted from $\pi N$ subthreshold 
coefficients by means of dispersion relations and hence may contain both 
experimental and theoretical uncertainties. 
This means that, in the case of the interactions given in Fig. \ref{fig2}, 
predictions from families I and II are very reliable whereas those 
associated with family III may be less so. For this reason it is important 
to establish how the results discussed in the preceding section depend on 
the various families of diagrams. 

In order to assess the importance of each family we show, in 
Fig. \ref{fig:ch6-contrib}, their relative contributions to the 
components of the TPEP. A general pattern one can observe is that 
two-loop contributions (family II) are 
negligible and, in particular, exactly zero for 
$V_{LS}^{+}$, $V_{SS}^{+}$, and $V_{T}^{+}$. 
The various 
profile functions are neatly dominated by either family I or III. 
The former, which is very precise, dominates the channels 
$V_{C}^{-}$, $V_{LS}^{+}$, $V_{SS}^{+}$, and $V_{T}^{+}$ 
and a modification on the values of the 
LECs would hardly influence the corresponding curves. 
This can be viewed as a strong constraint on the construction of 
phenomenological potentials. 
For the remaining channels, this condition is somehow relaxed, since 
they are dominated by the diagrams of family III. If one wishes, the 
freedom in these channels may be used to fix experimentally the LECs by 
means of $NN$ data. 


\begin{figure}[!tbp]
\begin{center}
\begin{tabular}{cc}
\epsfig{figure=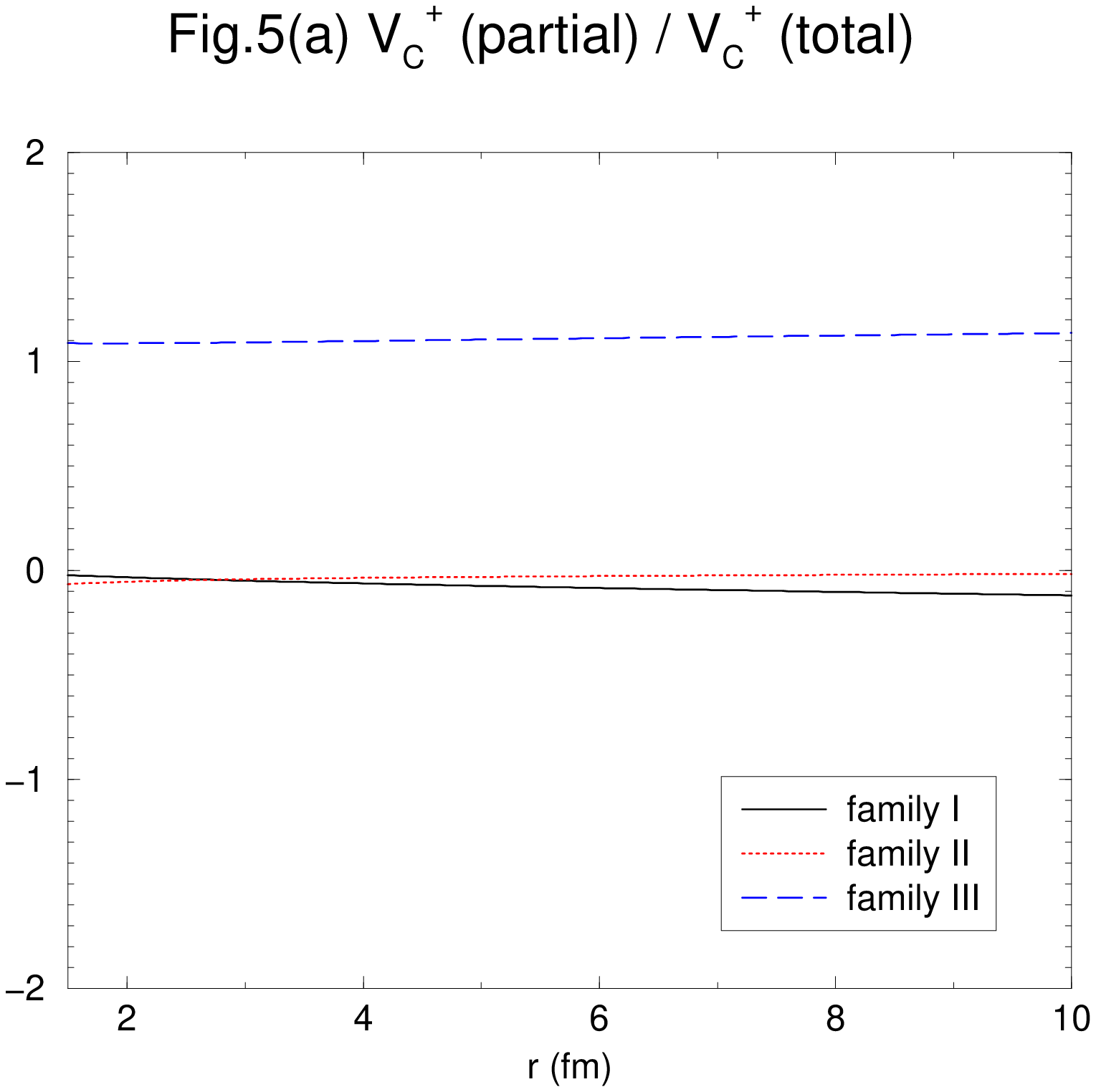, height=3in}&
\epsfig{figure=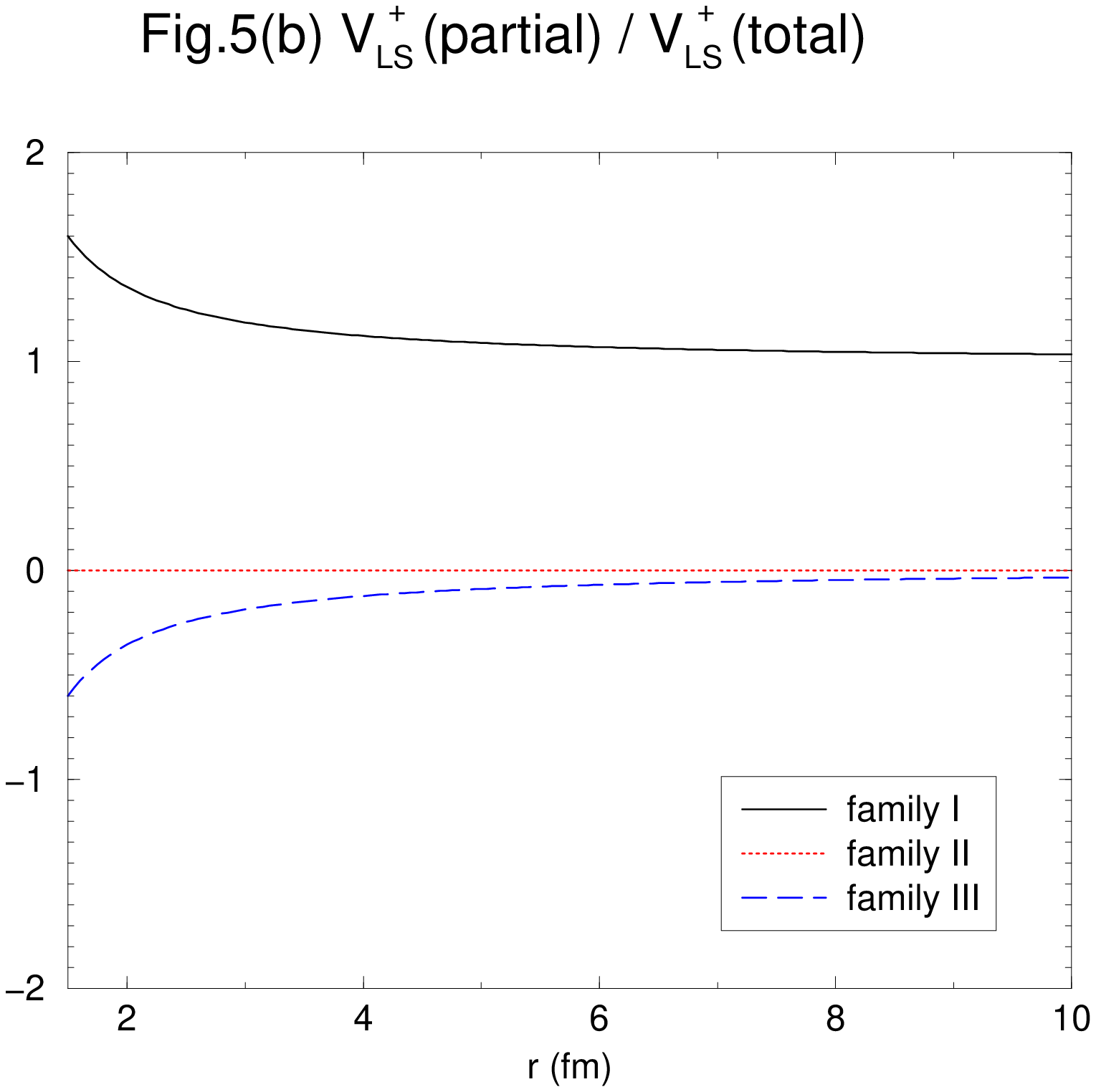, height=3in}\\[3mm]
\epsfig{figure=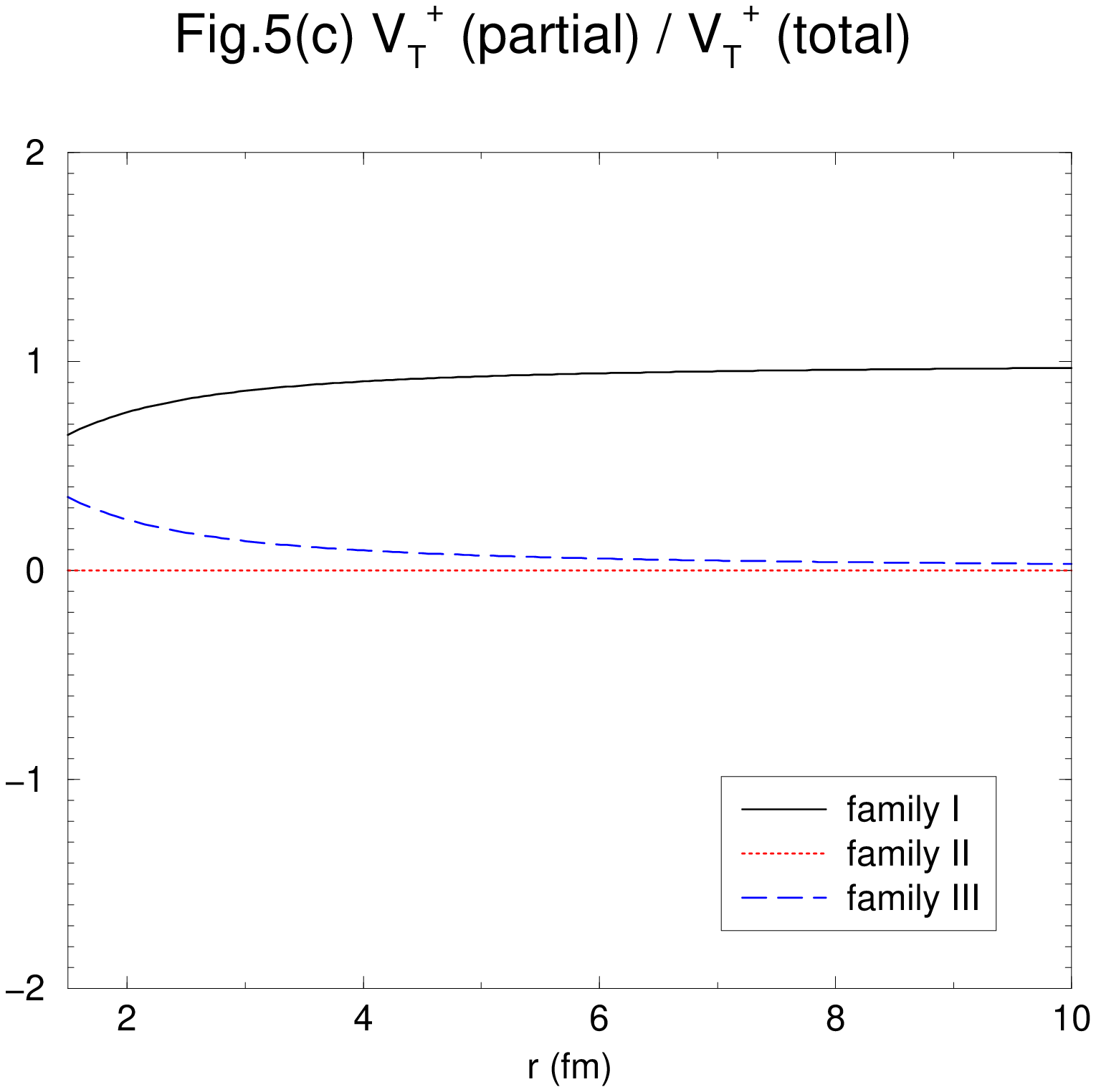, height=3in}&
\epsfig{figure=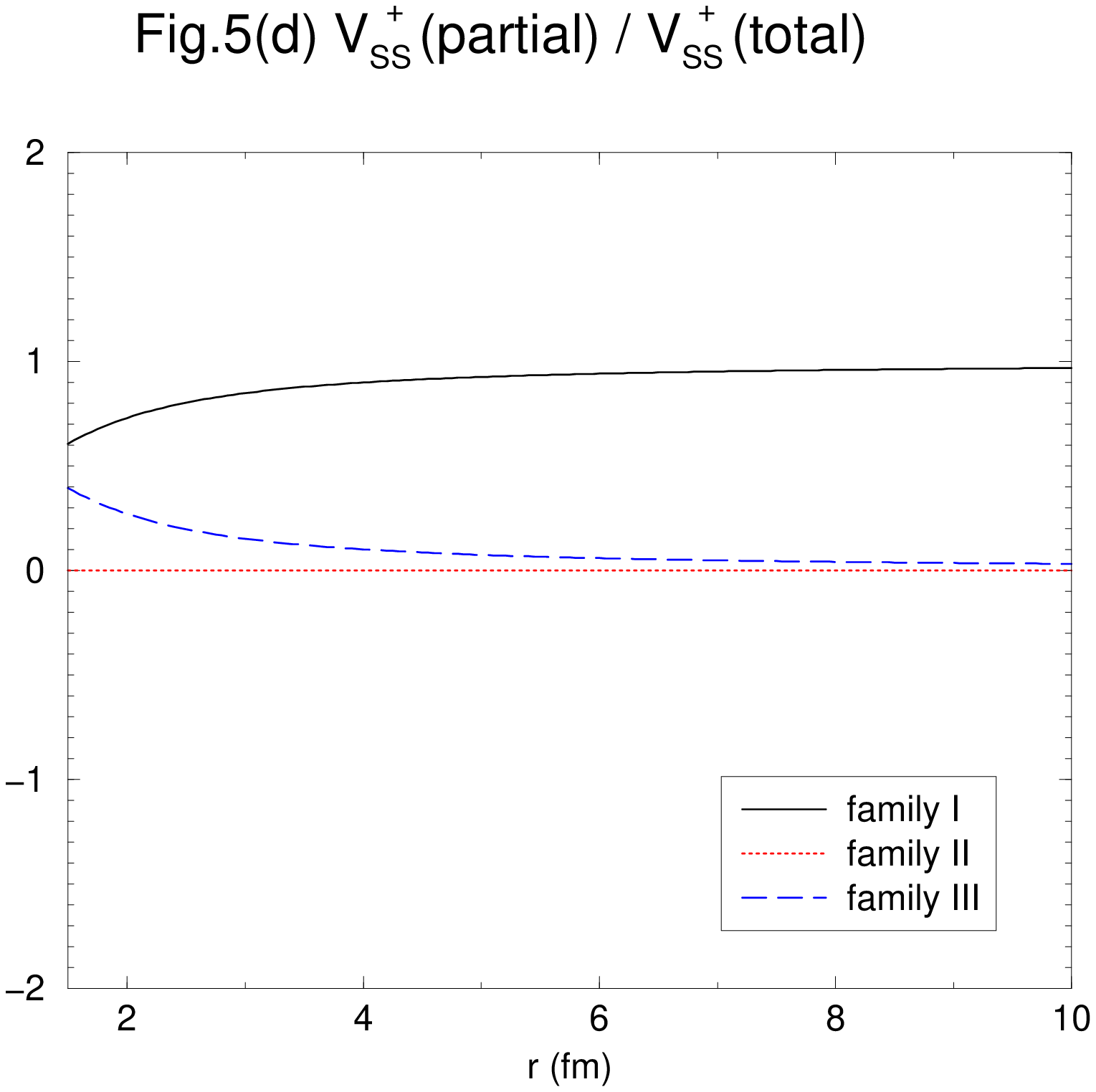, height=3in}\\[3mm]
\end{tabular}
\end{center}
\end{figure}

\newpage
\begin{figure}[!tbp]
\begin{center}
\begin{tabular}{cc}
\epsfig{figure=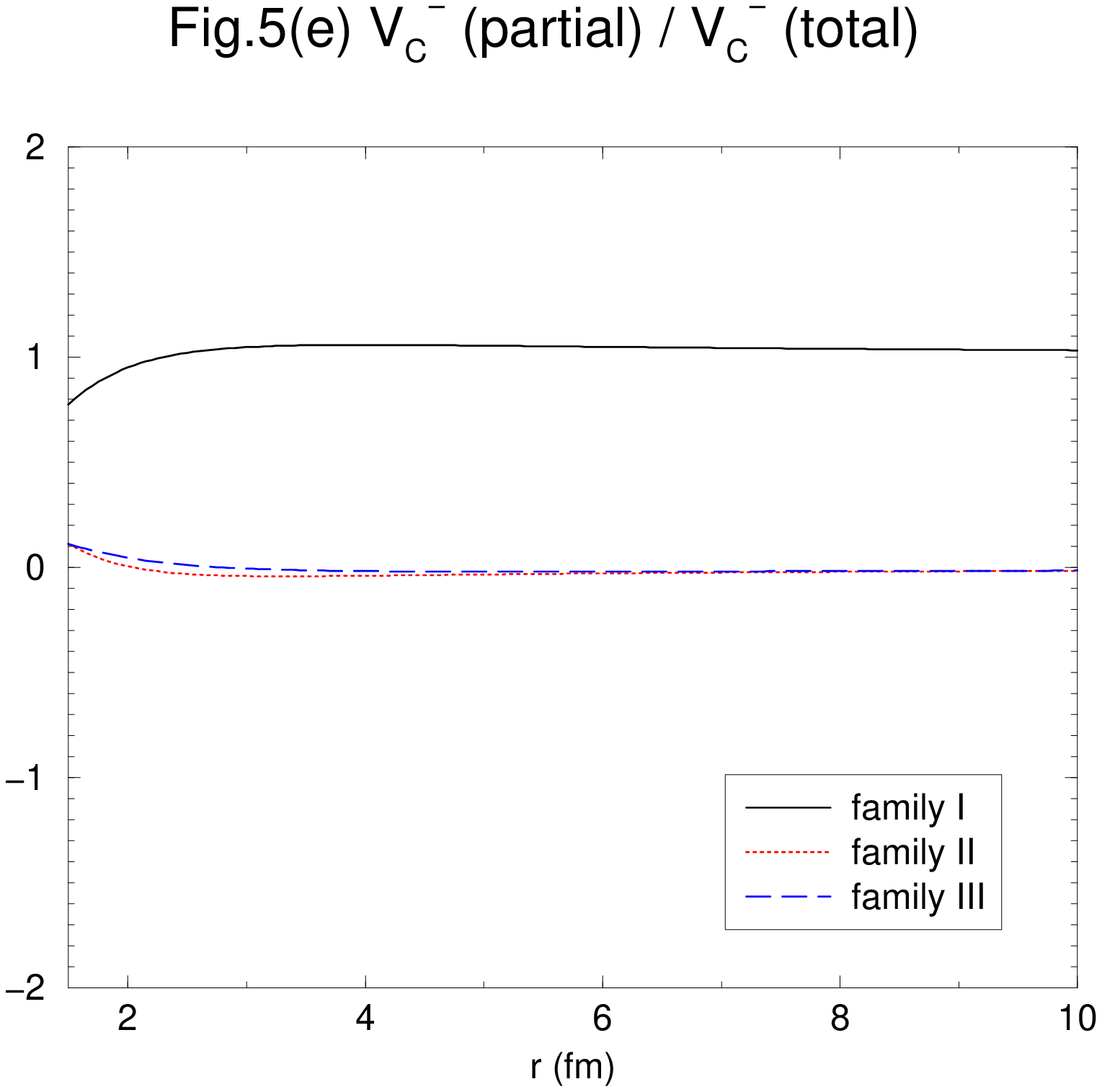, height=3in}&
\epsfig{figure=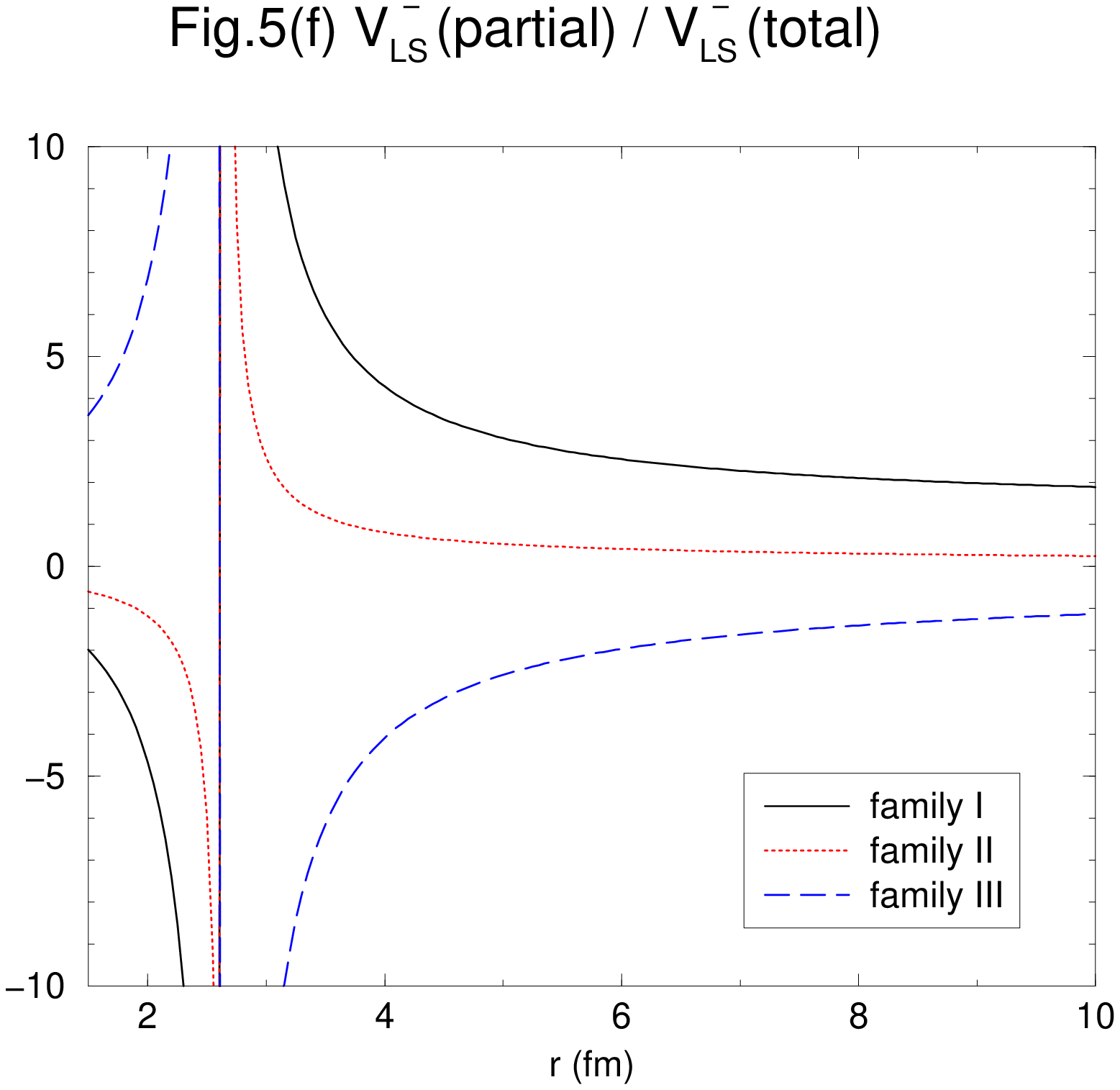, height=3in}\\[3mm]
\epsfig{figure=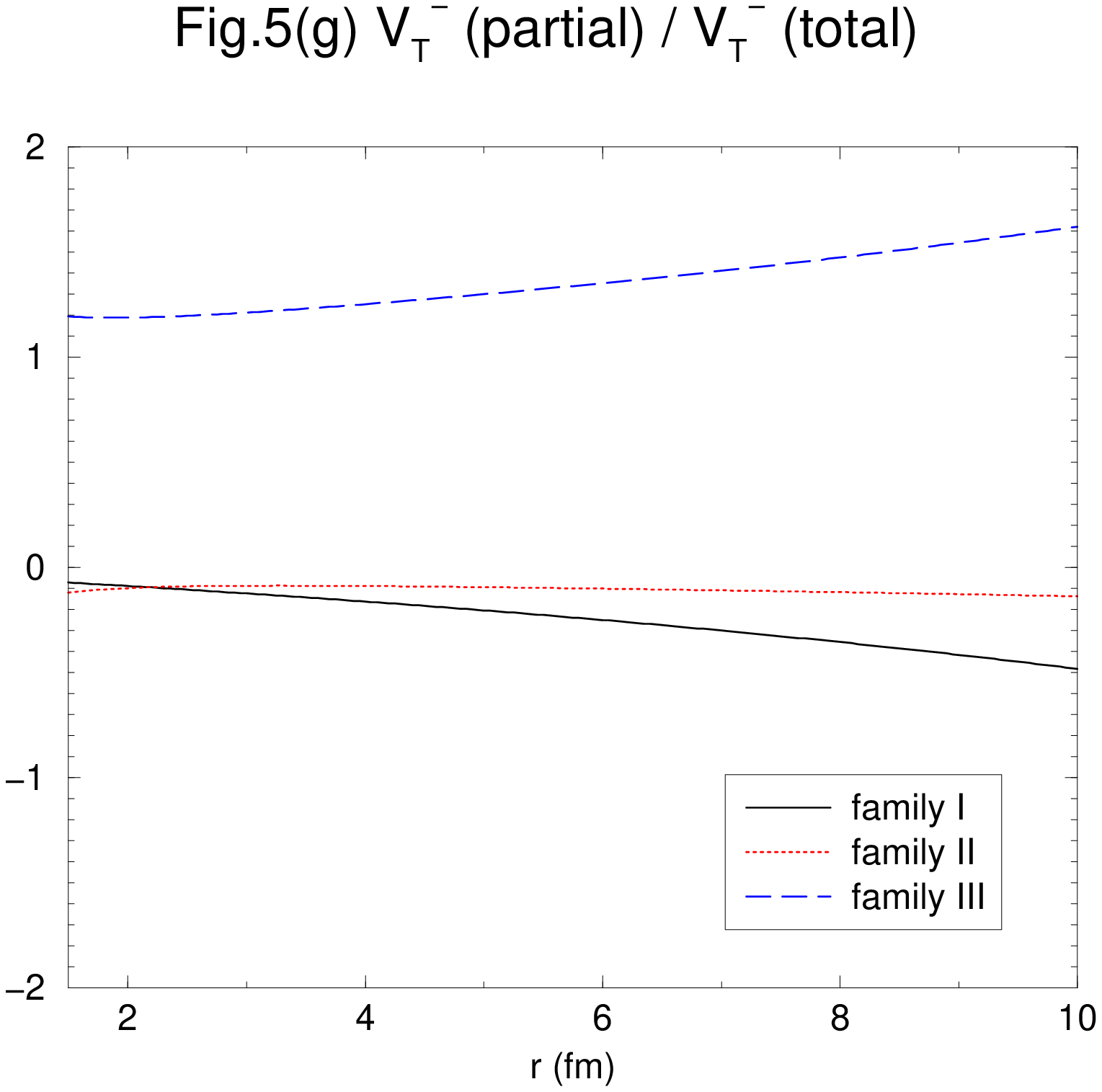, height=3in}&
\epsfig{figure=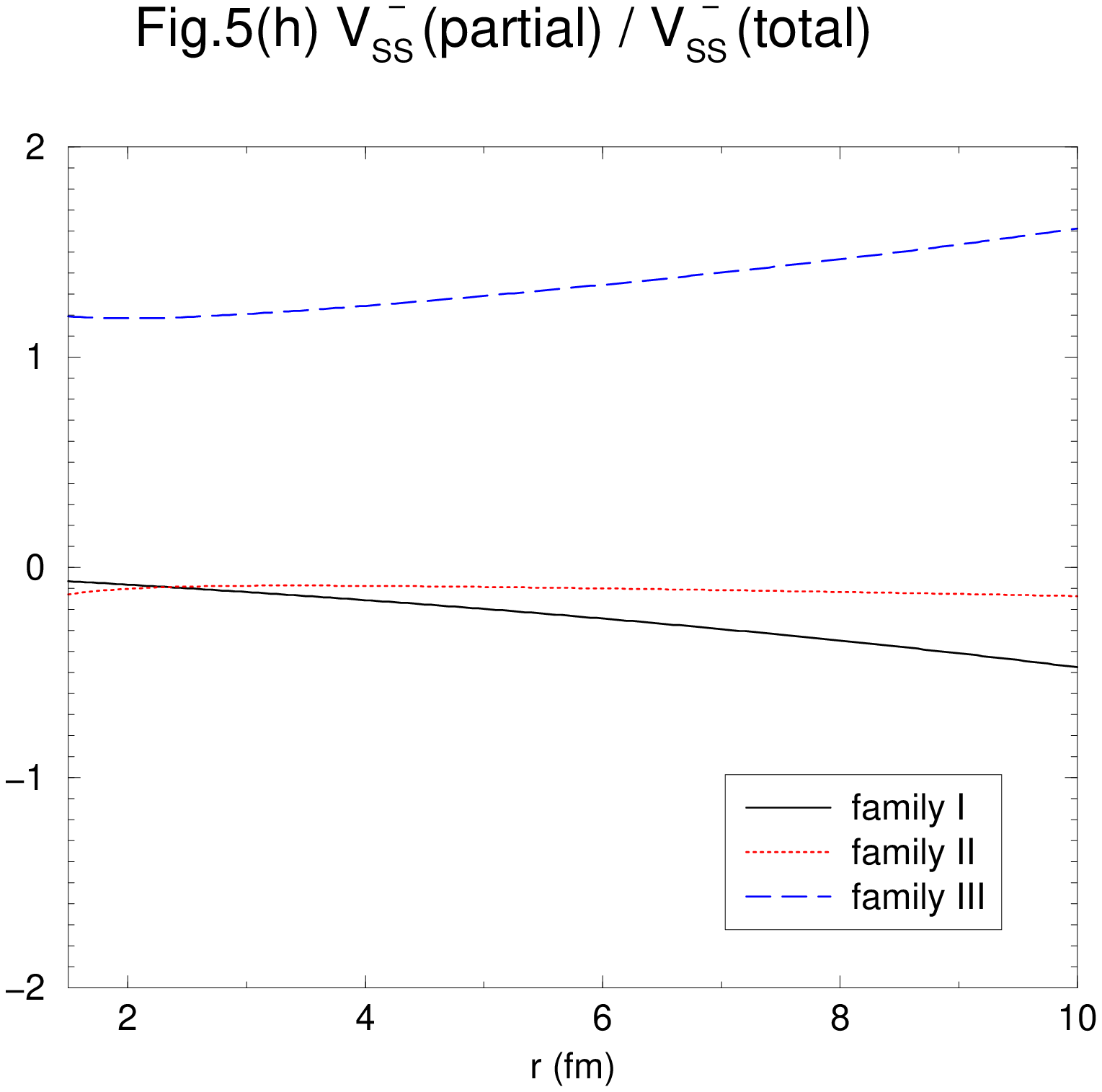, height=3in}
\end{tabular}
\end{center}
\caption{Relative weight of each family in the TPEP, obtained by 
dividing the partial contributions by the full result.}
\label{fig:ch6-contrib}
\end{figure}

\section{chiral structure} \label{secV}

In this section, we discuss chiral scales. In the case of the central 
components, these scales can be read directly from the functions $U^{\pm}_C$, 
given by Eqs. (\ref{3.9}) and (\ref{3.12}). For the other terms, 
there is a factor $(\mu^2/m^2)$ in the relation between $V^{\pm}$ and 
$U^{\pm}$, arising from the non-relativistic expansion of the Dirac 
spinors, (\ref{3.5})--(\ref{3.7}). Thus, in the $O(q^4)$ 
potential, one expands the corresponding functions $U^{\pm}$ up 
to $O(q^2)$. 

The leading term of the chiral TPEP is $O(q^2)$ and our results 
are written as sums of $O(q^2)$, $O(q^3)$, and $O(q^4)$ 
contributions. In the cases of $V_{T}^{+}$, $V_{SS}^{+}$, and $V_{C}^{-}$, 
this structure is mapped directly into the corresponding profile functions. 
The other components begin at $O(q^3)$. 

\begin{figure}[!tbp]
\begin{center}
\begin{tabular}{cc}
\epsfig{figure=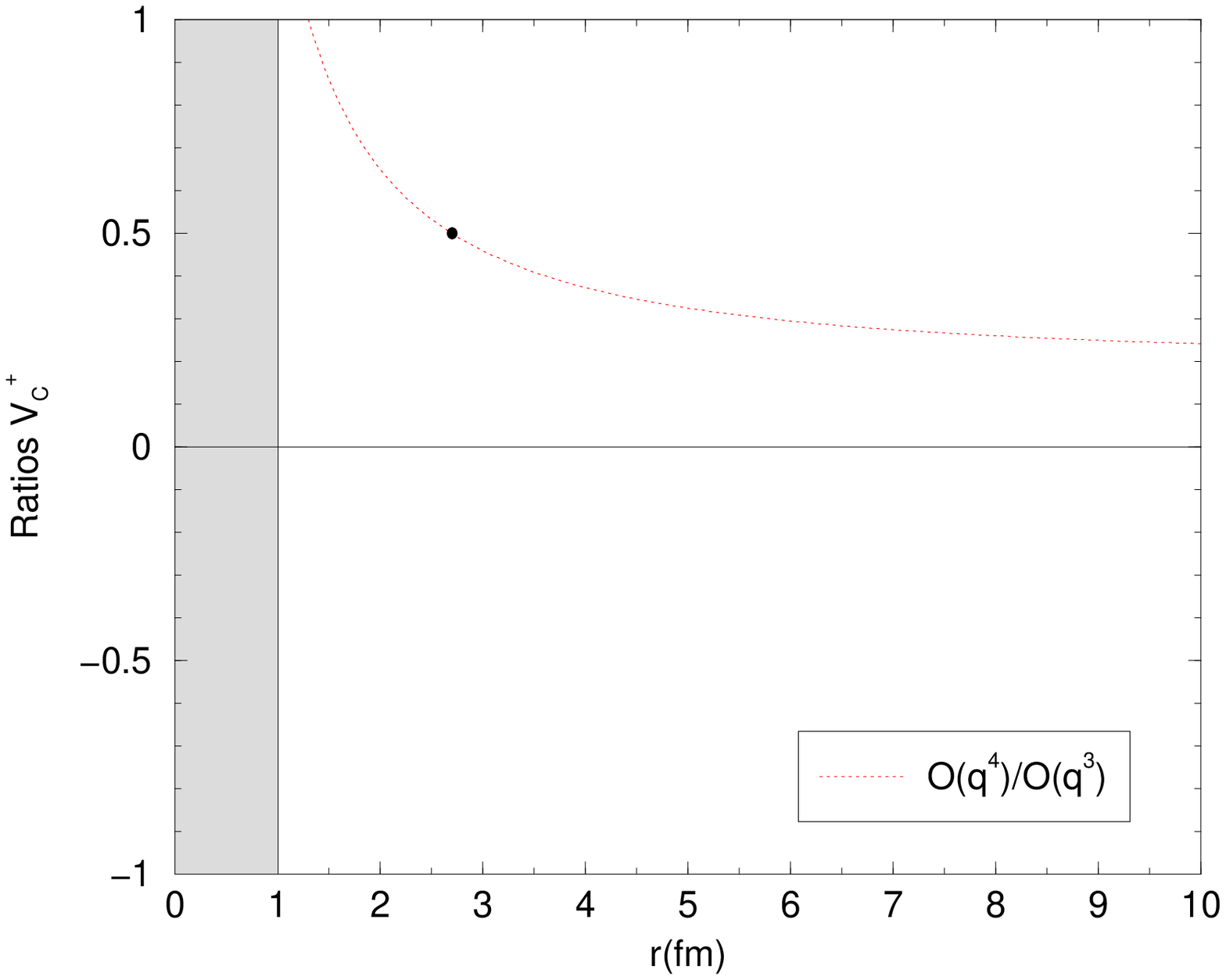, height=2.1in}&
\epsfig{figure=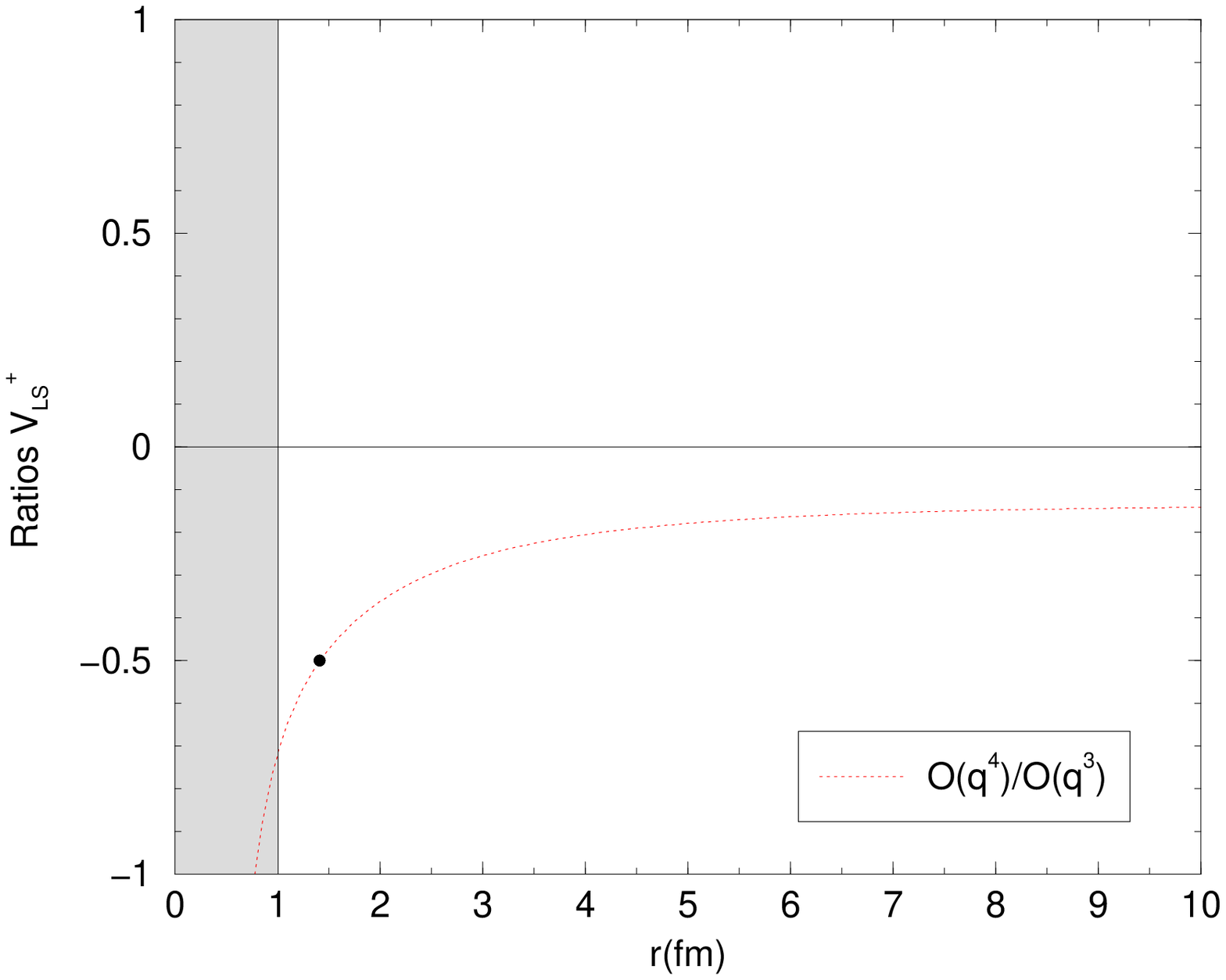, height=2.1in}\\
\epsfig{figure=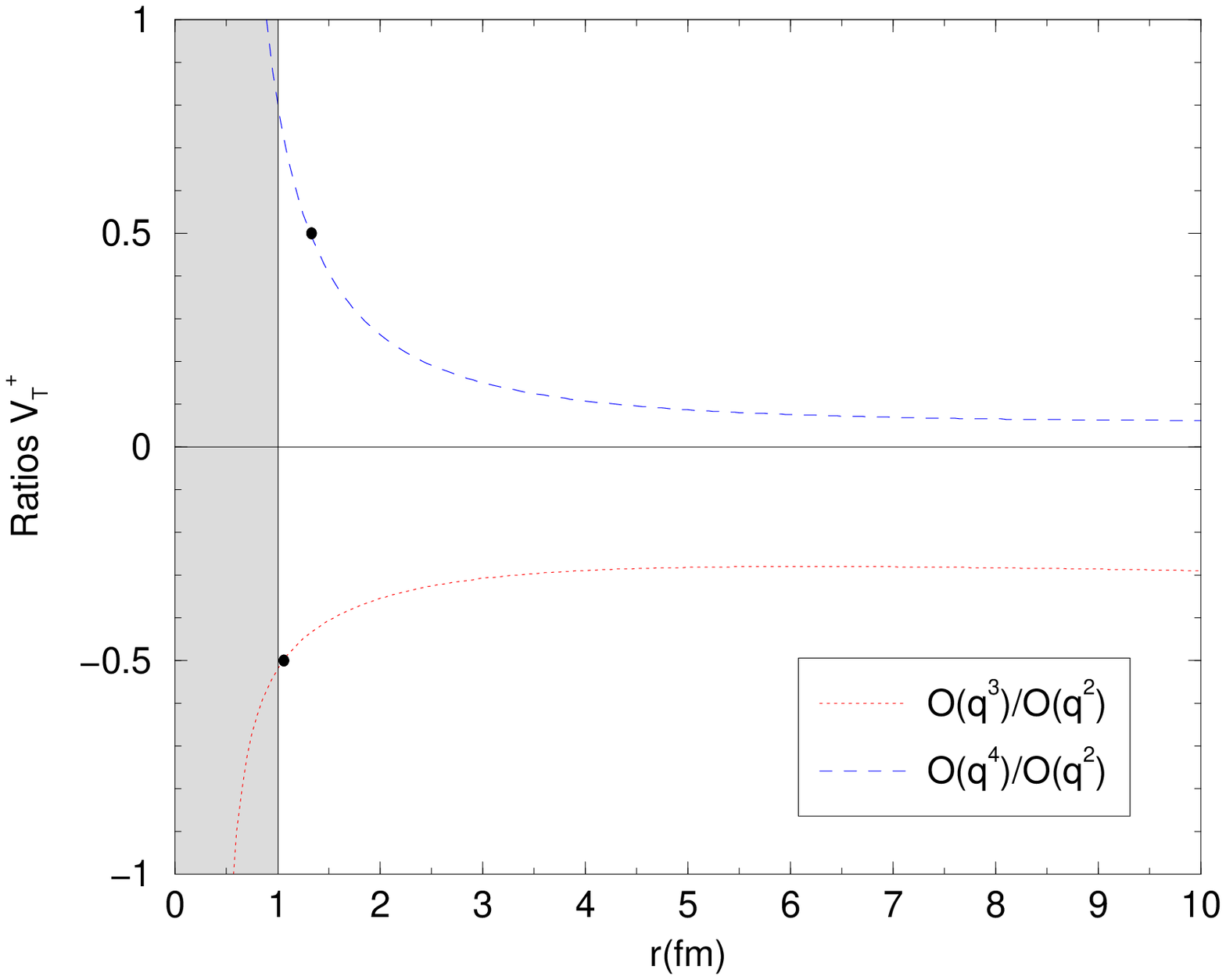, height=2.1in}&
\epsfig{figure=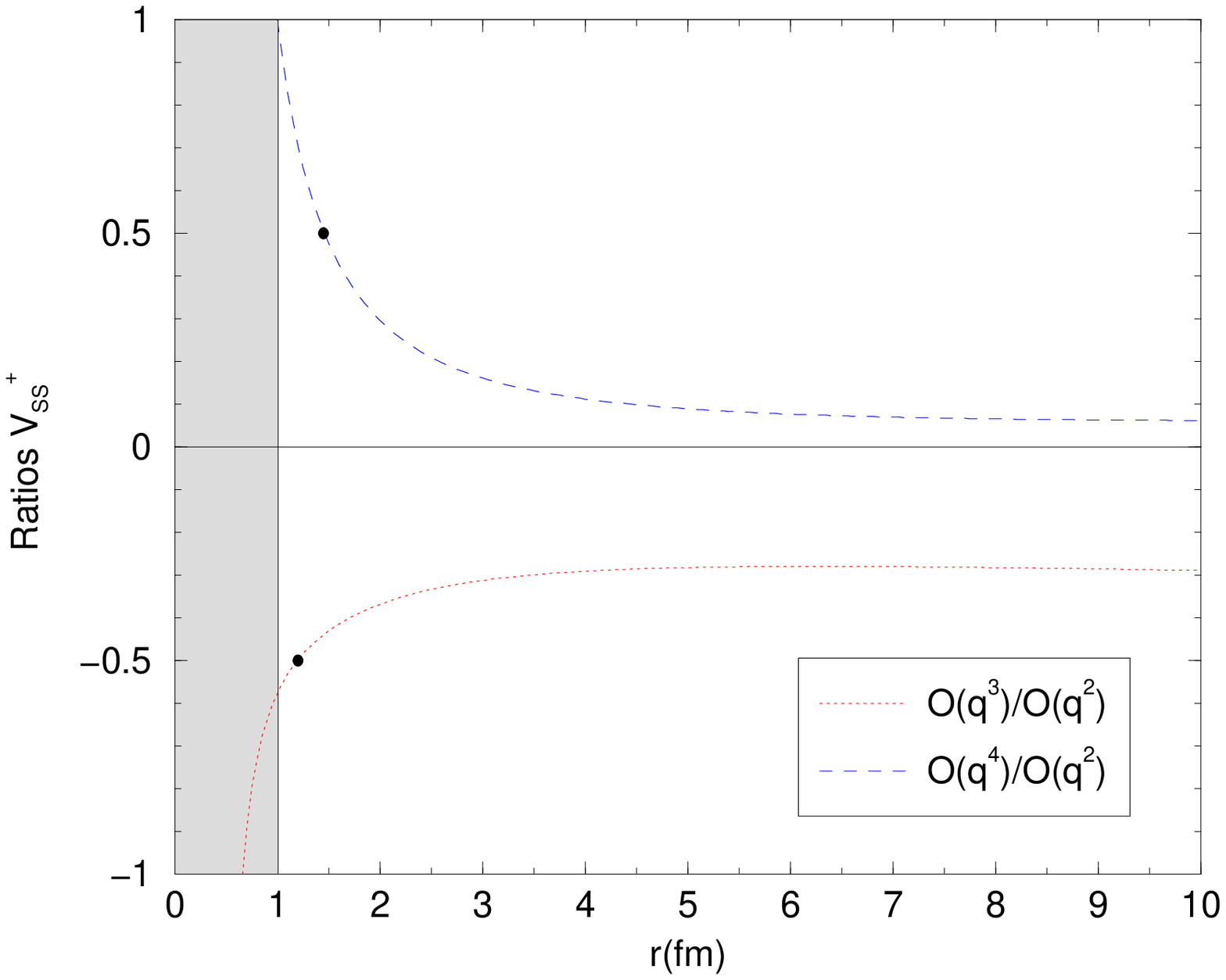, height=2.1in}\\
\epsfig{figure=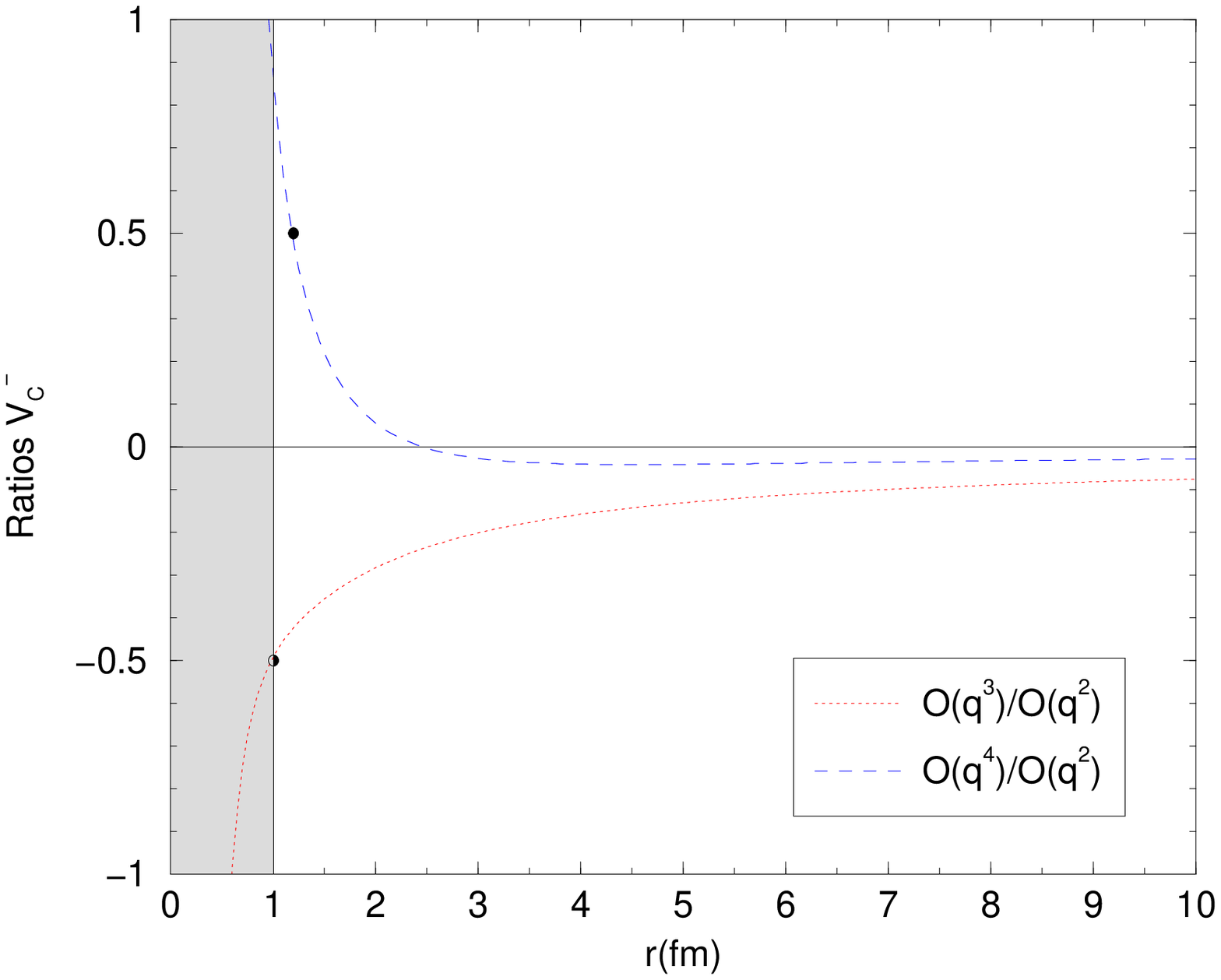, height=2.1in}&
\epsfig{figure=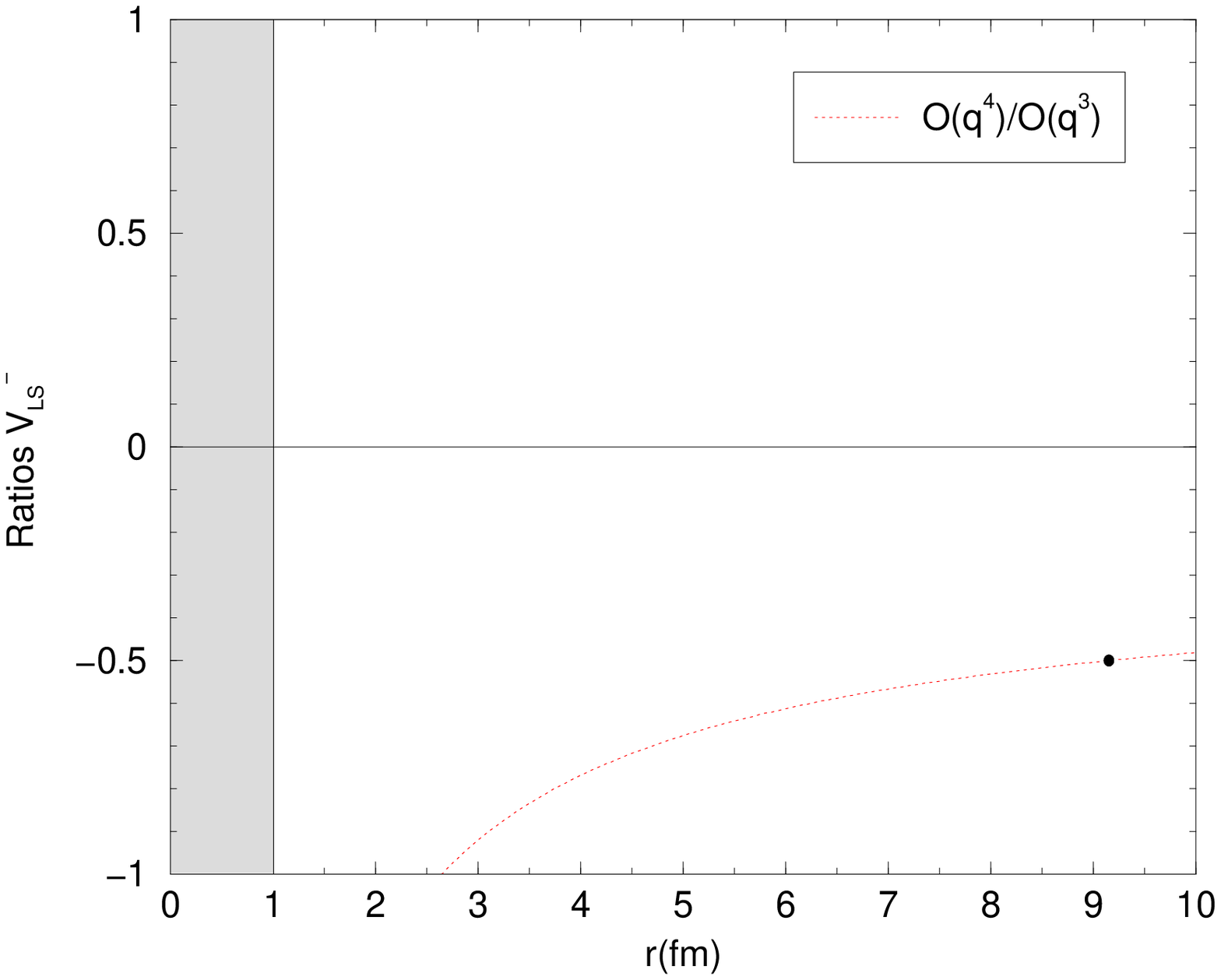, height=2.1in}\\
\epsfig{figure=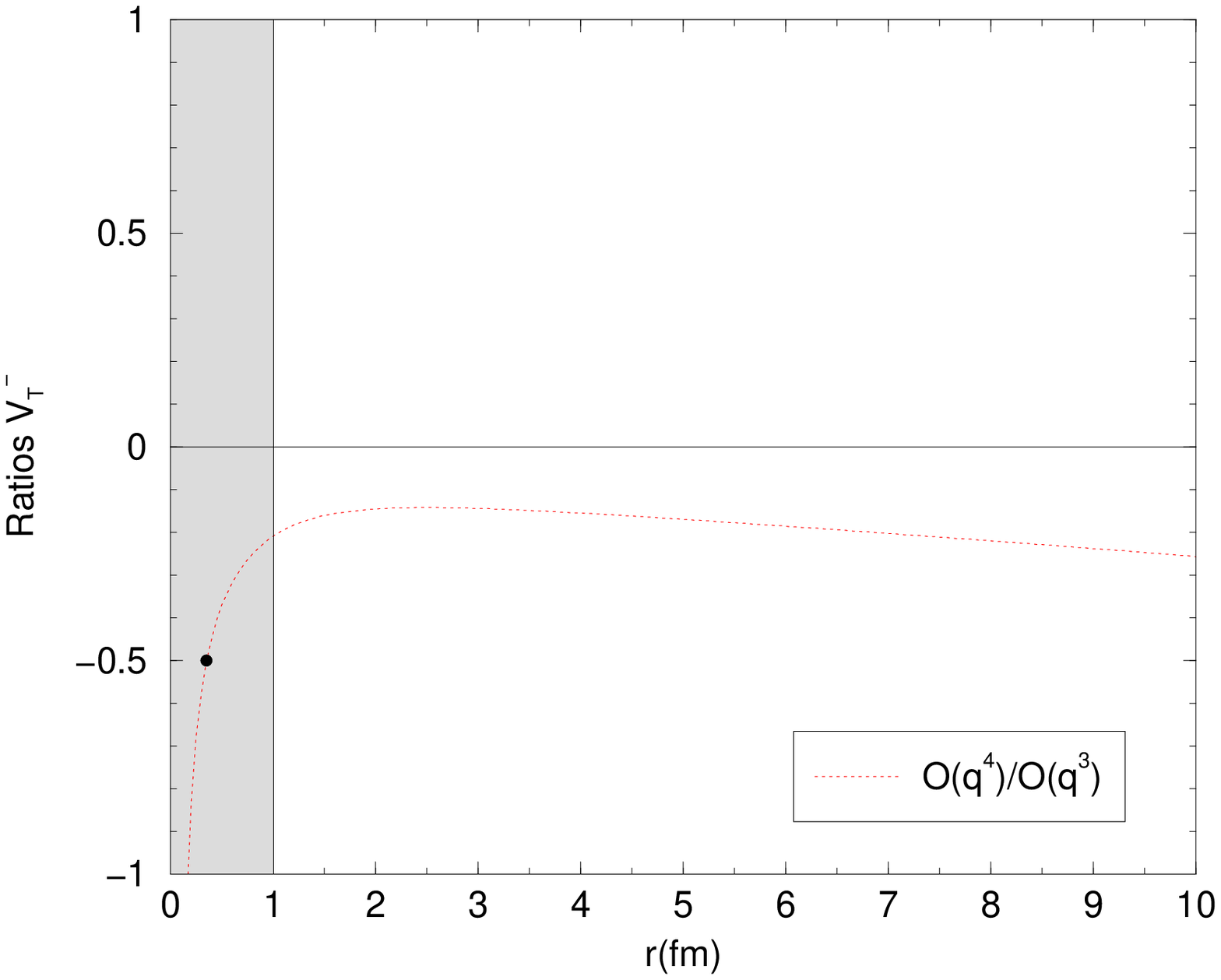, height=2.1in}&
\epsfig{figure=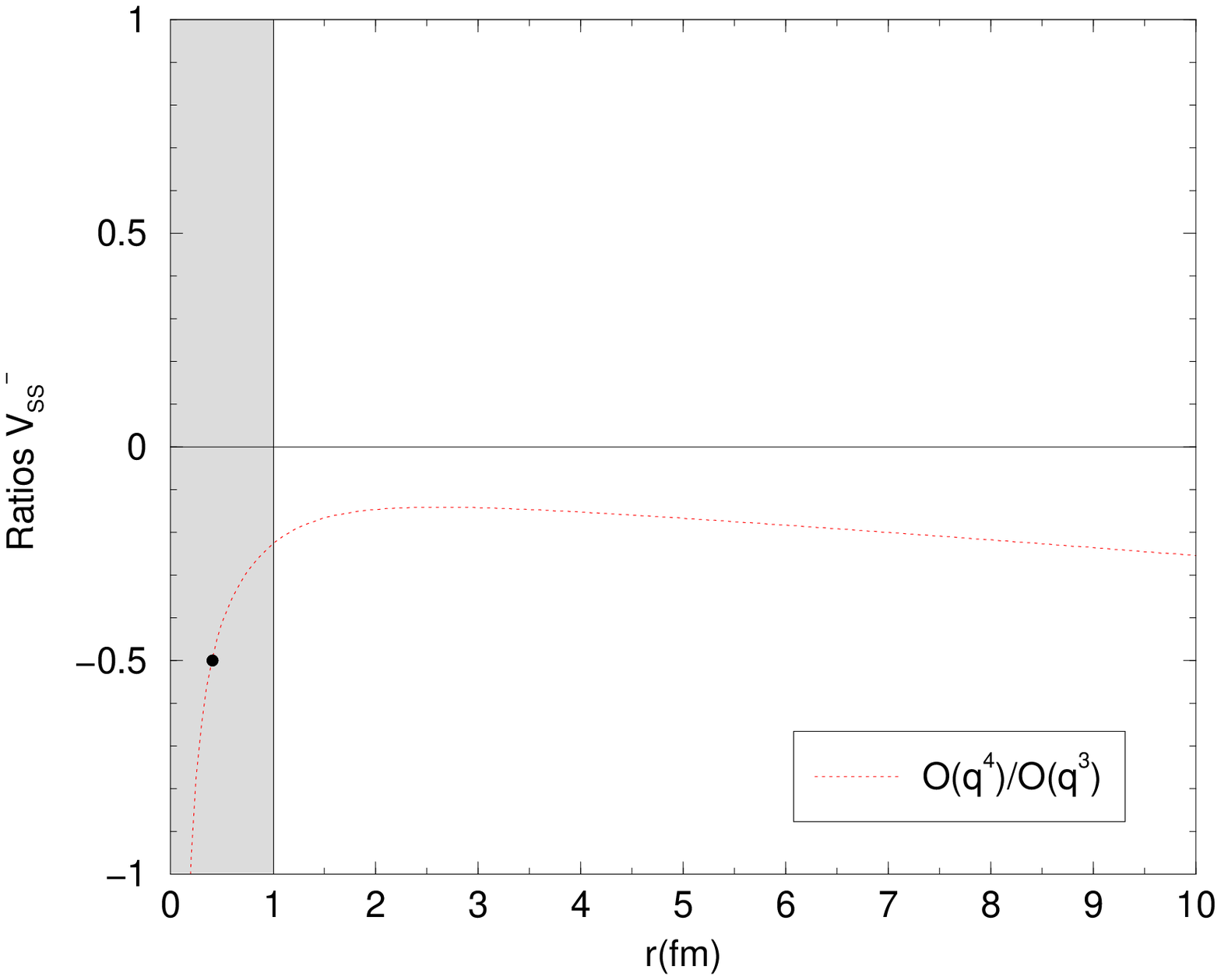, height=2.1in}
\end{tabular}
\end{center}
\caption{Relative contribution of each chiral order to the TPEP. The 
point in the curve where the ratio is 0.5 is indicated by a black dot, 
for the sake of guiding the eye.}
\label{chlayers}
\end{figure}

In $p$-space, the chiral series involves nucleon three-momenta, 
assumed to be small. This means that, in $r$-space, the chiral 
structure should become apparent at large distances. 
In order to check this, in Fig. \ref{chlayers} we show the ratios of 
the chiral layers for the various components of the potential. 
In all figures it is possible to note, at large distances, a rather 
well defined chiral hierarchy. Corrections are always smaller than the 
terms they correct. On the other hand, this hierarchy tends to break 
down when distances decrease and we assume that our results are not 
physical for $r<$ 1 fm. 
In two cases, namely, $V_{C}^{+}$ and $V_{LS}^{-}$, corrections are large 
within the region of physical interest. 


\section{the heavy baryon approximation} \label{secVI}

The relativistic potential is expressed by Eqs. (\ref{3.4})--(\ref{3.7}), 
(\ref{3.9})--(\ref{3.14}), and involves eight basic functions, 
denoted generically by $S(x)$. They are given by 
Eqs. (\ref{3.15})--(\ref{3.26}), and represent 
bubble, triangle, crossed box, planar box, double 
bubble, and double triangle diagrams. 
These functions have been derived by means of covariant techniques and 
correspond to the signature of relativity in this problem. Only the bubble 
integral can be evaluated analytically and the other ones are not 
homogeneous functions of either the pion mass or external three-momenta. 
In general, the expansion of the functions $S(x)$ 
in powers of $q/m$ is not mathematically defined. However, as 
discussed by Ellis and Tang \cite{EllisT}, if one forces such an 
expansion, one recovers 
{\em formally} the results of HBChPT. 
In Ref.\cite{HR} we have expanded our $O(q^4)$ $p$-space 
relativistic potential in this way and obtained (inequivalent) 
expressions that reproduce most of the standard $O(q^4)$ HBChPT 
results \cite{KBW,K3,K4,EM1}. Differences are due to the 
Goldberger-Treiman discrepancy and to the procedure adopted for 
subtracting the iterated OPEP. 
In this section, we discuss the numerical implications of the heavy 
baryon approximation in configuration space. 

We begin by considering the triangle integral $S_t$, given by 
Eq. (\ref{3.17}), that can also be expressed as \cite{BL1} 

\begin{equation}
S_t  = - \frac{8m}{\mu} \int_{4\mu^2}^\infty  dt'\; \mbox{Im}\gamma(t')\; 
\frac{e^{-\sqrt{t'/\mu^2} \,x}}{x}
\label{hba.1}
\end{equation}

\noindent
with 

\begin{equation}
\mbox{Im}\; \gamma(t') = \frac{1}{8\pi \sqrt{t'(4m^2-t')}}
\,\arctan \left[  \frac{\sqrt{(4m^2-t')(t'-4\mu^2)}}{t'-2\mu^2}\right]\,.
\label{hba.2}
\end{equation}

The heavy baryon approximation consists in writing 

\begin{equation}
\mbox{Im}\;\gamma(t') \simeq \frac{1}{16\pi m\sqrt{t'}}
\,\arctan \left[ \frac{2m\sqrt{t'-4\mu^2}}{t'-2\mu^2}\right]
\label{hba.3}
\end{equation}

\noindent 
and treating formally the argument 

\begin{equation}
\alpha = \frac{2m\sqrt{t'-4\mu^2}}{t'-2\mu^2}
\label{hba.4}
\end{equation}

\noindent
as being $O(q^{-1})$.
This would suggest that one could use the result 
$\arctan \alpha = \pi/2 - 1/\alpha + 1/3\alpha^3+\cdots $ 
in order to derive the heavy baryon expansion of the triangle integral. 
Recently BL \cite{BL1} have discussed the 
properties of the spectral representation based on Eq. (\ref{hba.4}) 
and remarked that the series for $\arctan $ which underlies the heavy 
baryon approximation is valid only in the domain $|\alpha|\geq 1$. For 
$|\alpha|<1$ one should use $\arctan \alpha = \alpha - \alpha^3/3+\cdots$, 
but this corresponds to an expansion in {\em inverse} powers of $q$. 
They showed\footnote{They worked in momentum space.} that a suitable 
representation for $S_t$ is 

\begin{eqnarray}
S_t^{B\!L}&=&-\frac{8m}{\mu}\int_{4\mu^2}^\infty dt'\;
\frac{e^{-\sqrt{t'/\mu^2} \,x}}{x}\;\frac{1}{16\pi m\sqrt{t'}}
\,\arctan \left[ \frac{2m\sqrt{t'-4\mu^2}}{t'-2\mu^2}\right]
\label{hba.a5}\\[2mm]&\approx&
- \frac{1}{2\pi m\mu}\int_{4\mu^2}^{\infty} dt'\;
\frac{1}{\sqrt{t'}} \left\{ \left[\frac{\pi}{2} - 
\frac{(t'-2\mu^2)}{2 m \sqrt{t'\!-\!4\mu^2}}\right]_{\!HB} 
\right.
\nonumber\\[2mm]
&+& \left. \left[ \frac{\mu \sqrt{t'}}{2 m \sqrt{t'\!-\!4\mu^2}}
-\frac{\sqrt{t'}}{2 \mu}\arctan \frac{\mu^2}{m\sqrt{t'\!-\!4\mu^2}}
\right]_{\!t\!h}\right\}\; \frac{e^{-\sqrt{t'/\mu^2} \,x}}{x}\;. 
\label{hba.5}
\end{eqnarray}

The heavy baryon approximation consists in keeping only the first 
bracket in 
the integrand. However, this does not cover the region $t' \sim 4\mu^2$, 
where the second term dominates. As a consequence, the heavy baryon 
approximation of $S_t$, which reads 

\begin{equation}
S_t\rightarrow S_t^{H\!B} = \left[-\;\frac{e^{-2x}}{2 x^2}\right]^{LO}
+\left[\frac{\mu}{2m}\frac{2}{\pi x^2}\;
[xK_0(2x)+K_1(2x)]+\right]^{NLO}\cdots\;,
\label{hba.6}
\end{equation}

\noindent
is not suitable for all values of $x$, as observed numerically in our 
previous work \cite{R01}. The exponential in the integrand of 
Eq. (\ref{hba.1}) shows clearly that, for large values of $x$, results 
are dominated by the lower end of the integration. Thus, a good 
description of $S_t$ at large distances requires a decent representation 
for $\mbox{Im}\;\gamma(t')$ near $t'=4\mu^2$. 

\begin{figure}[!htb]
\begin{center}
\epsfig{figure=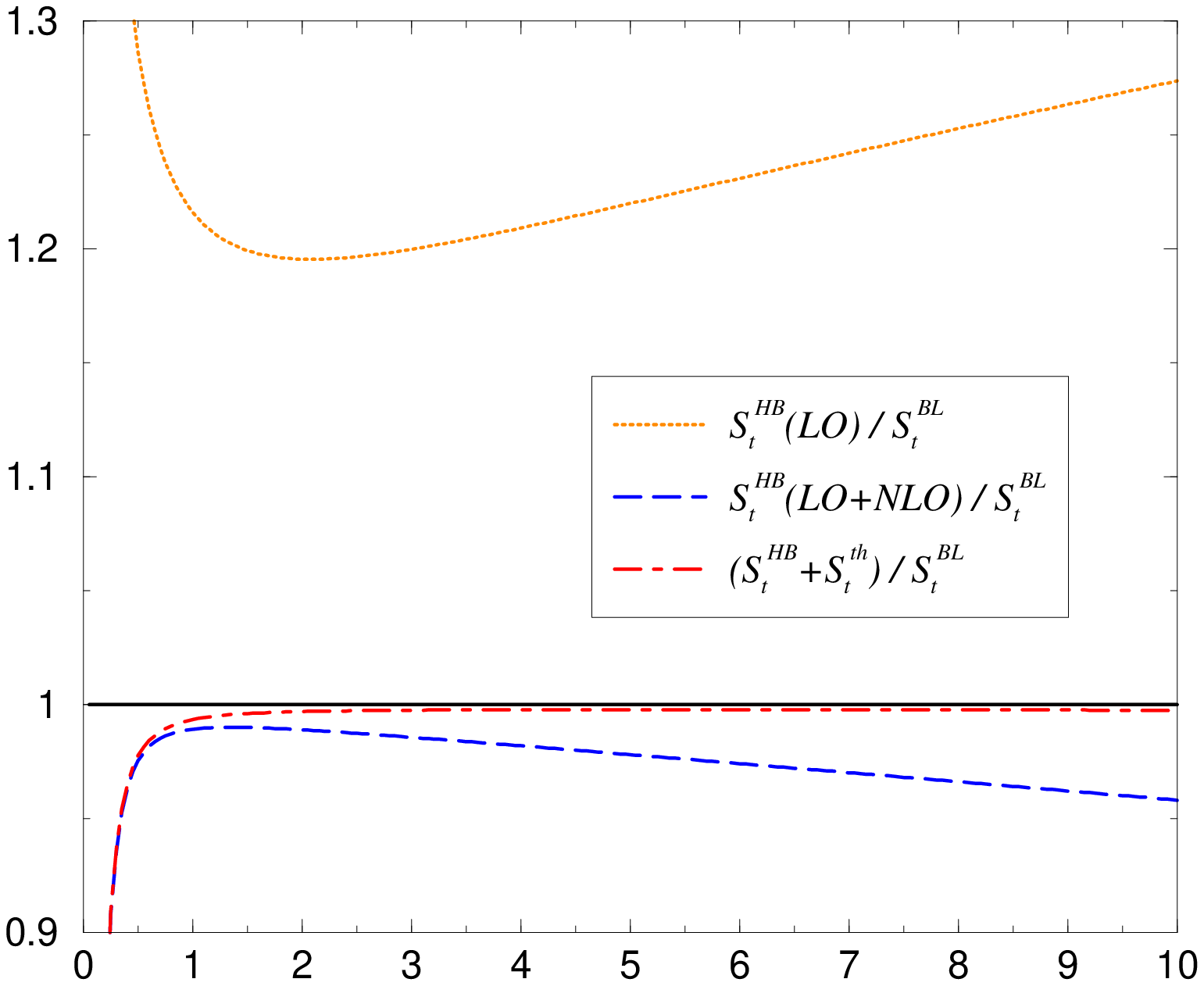, width=9cm}
\end{center}
\caption{The heavy baryon expansion of the triangle integral, given 
by Eq. (\ref{hba.6}), and the relativistic BL correction 
($S_t^{th}$), divided by $S_t^{BL}$.}
\label{fig:Pi_t_BL}
\end{figure}

In Fig. \ref{fig:Pi_t_BL} we display the ratios of the various terms 
of Eq. (\ref{hba.5}) by Eq. (\ref{hba.a5}). Inspecting this figure 
one learns that the first two terms of the heavy baryon series do not 
represent well the full result. In order to have a good description 
of $S_t^{BL}$ at large distances one has to add $S_t^{th}$ which, 
as pointed out by BL, cannot be obtained through the heavy baryon series. 

An advantage of the heavy baryon formalism is that it gives rise to 
power counting, which is absent in relativistic baryon ChPT based on 
dimensional regularization \cite{GSS}. In order to overcome this 
difficulty, BL proposed a new regularization 
scheme, based on a previous work by Ellis and Tang~\cite{EllisT}. The so 
called Infrared Regularization (IR) respects the correct analytic 
structure around the point $t'=4\mu^2$, is manifestly Lorentz invariant, 
and gives rise to power counting. 

In the case of the triangle integral, the infrared regularized 
expressions reads 

\begin{eqnarray}
S_t^{I\!R}&=&-\int_0^1da\int_0^\infty db\;
\frac{(1-b)\;2m/\mu}{\Lambda_t^2}\;\frac{e^{-\theta_t\,x}}{4\pi x}
\label{hba.7}
\end{eqnarray}

\noindent with $\Lambda_t^2$ and $\theta_t$ given by 
Eq. (\ref{3.17}). 

\begin{figure}[!htb]
\begin{center}
\epsfig{figure=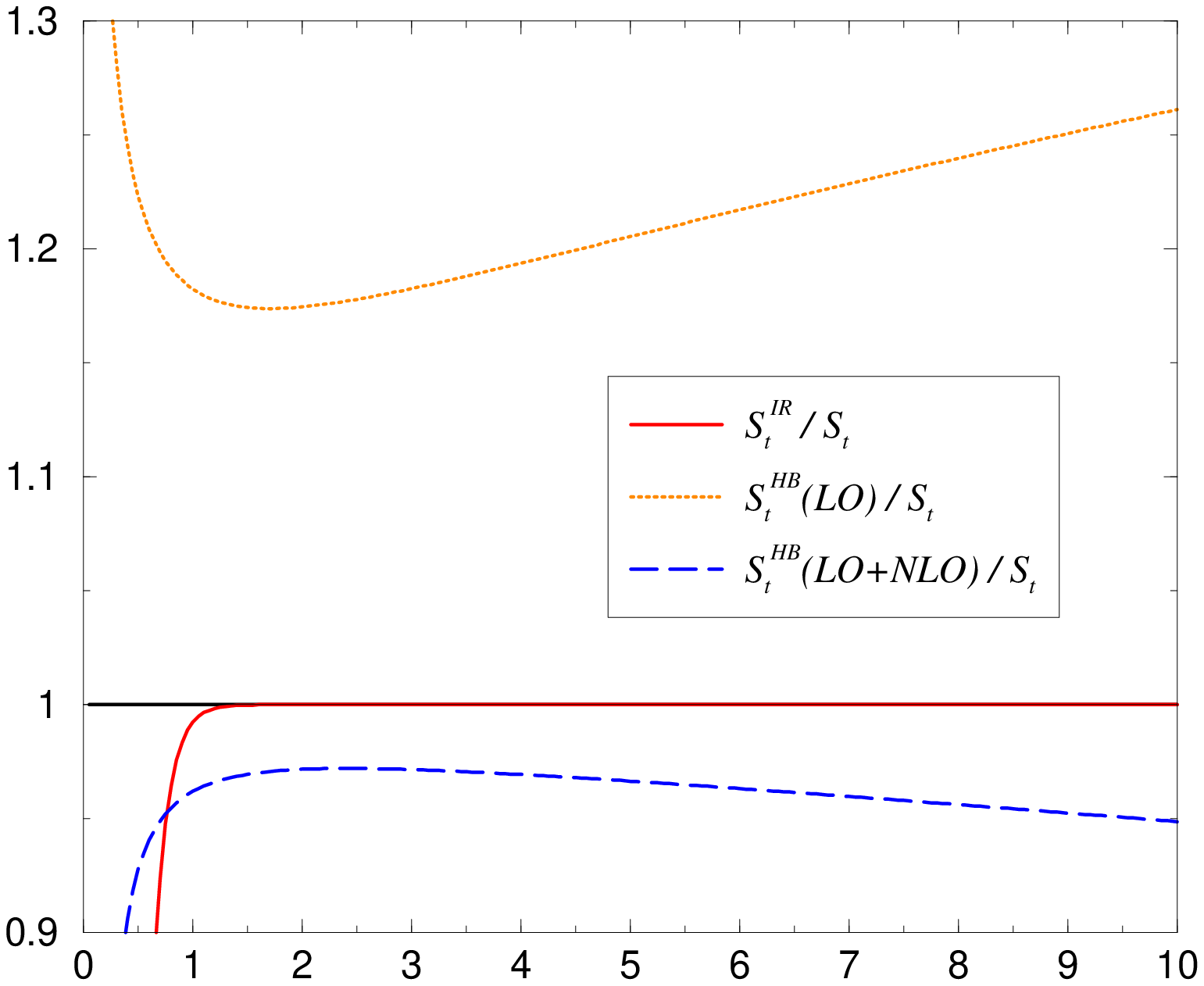, width=9cm}
\end{center}
\caption{The heavy baryon approximation of the triangle integral.}
\label{fig:Pi_t}
\end{figure}

In Fig. \ref{fig:Pi_t} we compare the infrared regularized triangle 
integral ($S_t^{I\!R}$) with that given by Eq. (\ref{3.17}), obtained 
through dimensional regularization. For comparison, we also plot the 
results of the heavy baryon formulation at $LO$ and $NLO$. The relativistic 
versions of the triangle integral are numerically identical for 
$r>$ 1.5 fm, indicating that the form of the regularization procedure 
is irrelevant in the region of physical interest. 

The discussion about the triangle integral may be extended to the 
functions $S_\times$, $S_b$, and $\tilde{S}_b$, associated 
with {\em crossed box} and {\em planar box} diagrams, and given by 
Eqs. (\ref{3.20})--(\ref{3.24}). The heavy baryon approximations for 
these results read \cite{HR} 

\begin{eqnarray}
S_\times^{H\!B}&\cong&\frac{2}{\pi x}\,K_0(2x)\!-\!\left[\frac{\mu}{m}\right]
\frac{e^{-2x}}{2x}\!-\!\left[\frac{\mu}{m}\right]^2\left\{\frac{1}{\pi x^2}
\left[2xK_0(2x)+(1\!-\!x^2)K_1(2x)\right]\right\}\!+\!\cdots\;,
\label{hba.8}\\[2mm]
S_b^{H\!B}&\cong&\frac{2}{\pi x}\,K_0(2x)\!-\!\left[\frac{\mu}{m}\right]
\frac{e^{-2x}}{4x}\!-\!\left[\frac{\mu}{m}\right]^2\left\{\frac{1}{3\pi x^2}
\left[2xK_0(2x)+(1\!-\!x^2)K_1(2x)\right]\right\}\!+\!\cdots\;,
\label{hba.9}\\[2mm]
\tilde{S}_b^{H\!B} &\cong& \frac{e^{-2x}}{4 x^2}
\!-\!\left[\frac{\mu}{m}\right]\frac{2}{3\pi x^2}\left[xK_0(2x)+K_1(2x)\right]
\!+\!\left[\frac{\mu}{m}\right]^2\frac{3\;e^{-2x}}{32x}\!+\!\cdots\;.
\label{hba.10}
\end{eqnarray}

\begin{figure}[!htb]
\begin{center}
\epsfig{figure=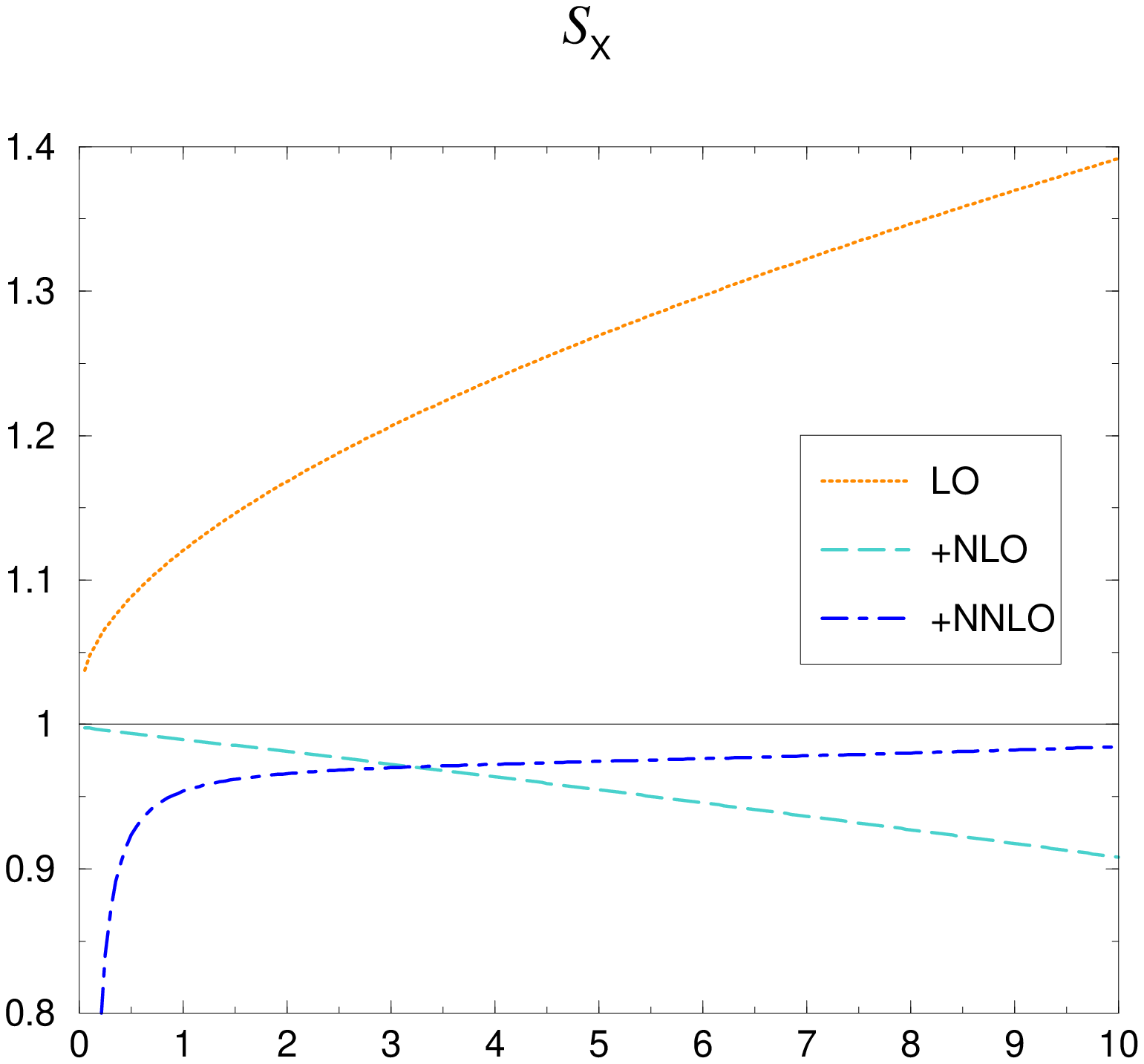, width=9cm}
\end{center}
\caption{The heavy baryon approximation of $S_{\times}$. The partial 
sums are divided by the relativistic result, Eq. (\ref{3.20}).}
\label{fig:Pi_x}
\end{figure}

\begin{figure}[!htb]
\begin{center}
\epsfig{figure=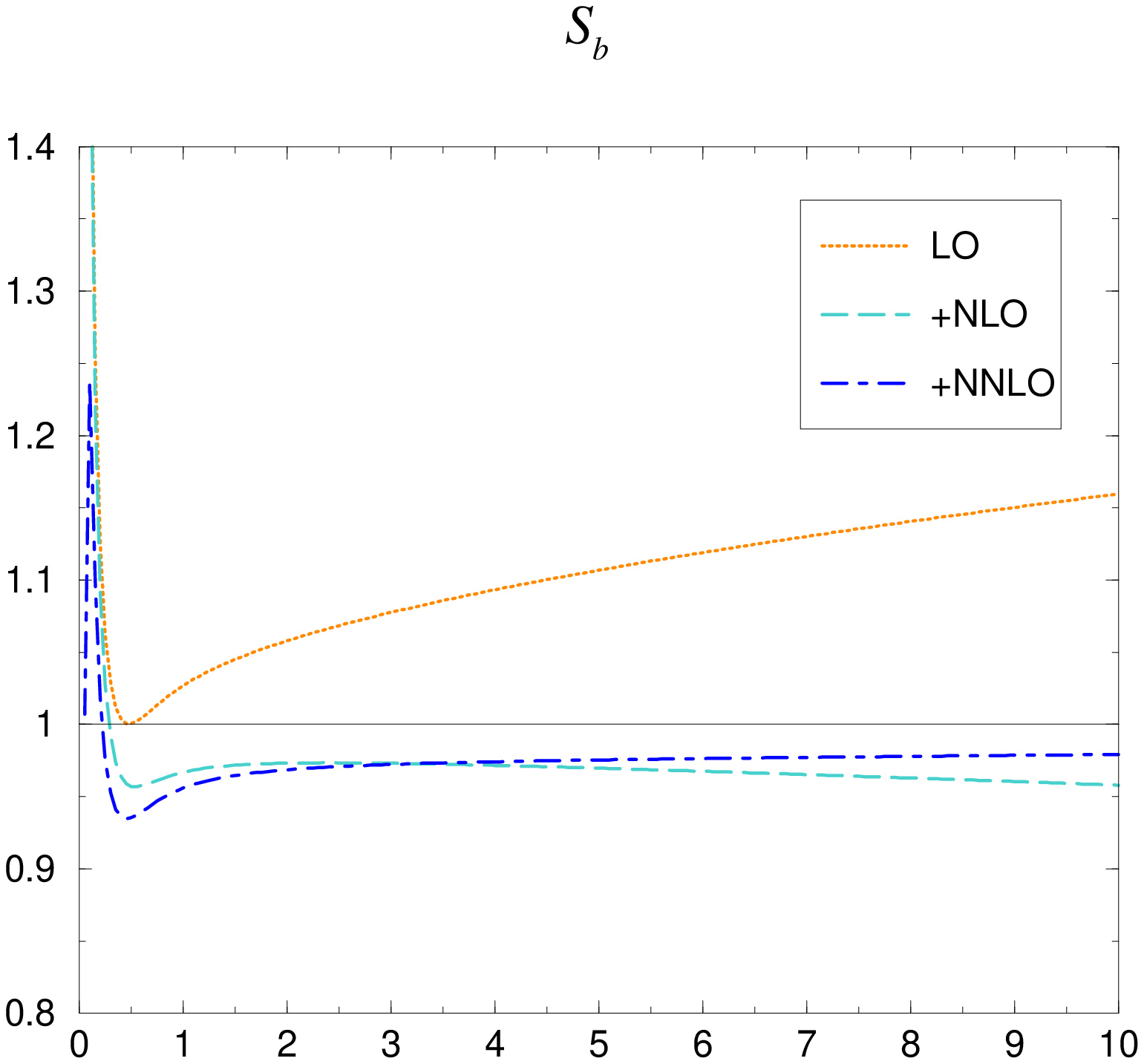, width=9cm}
\end{center}
\caption{The heavy baryon approximation of $S_b$. The partial 
sums are divided by the relativistic result, Eq. (\ref{3.21}).}
\label{fig:Pi_b}
\end{figure}

\begin{figure}[!htb]
\begin{center}
\epsfig{figure=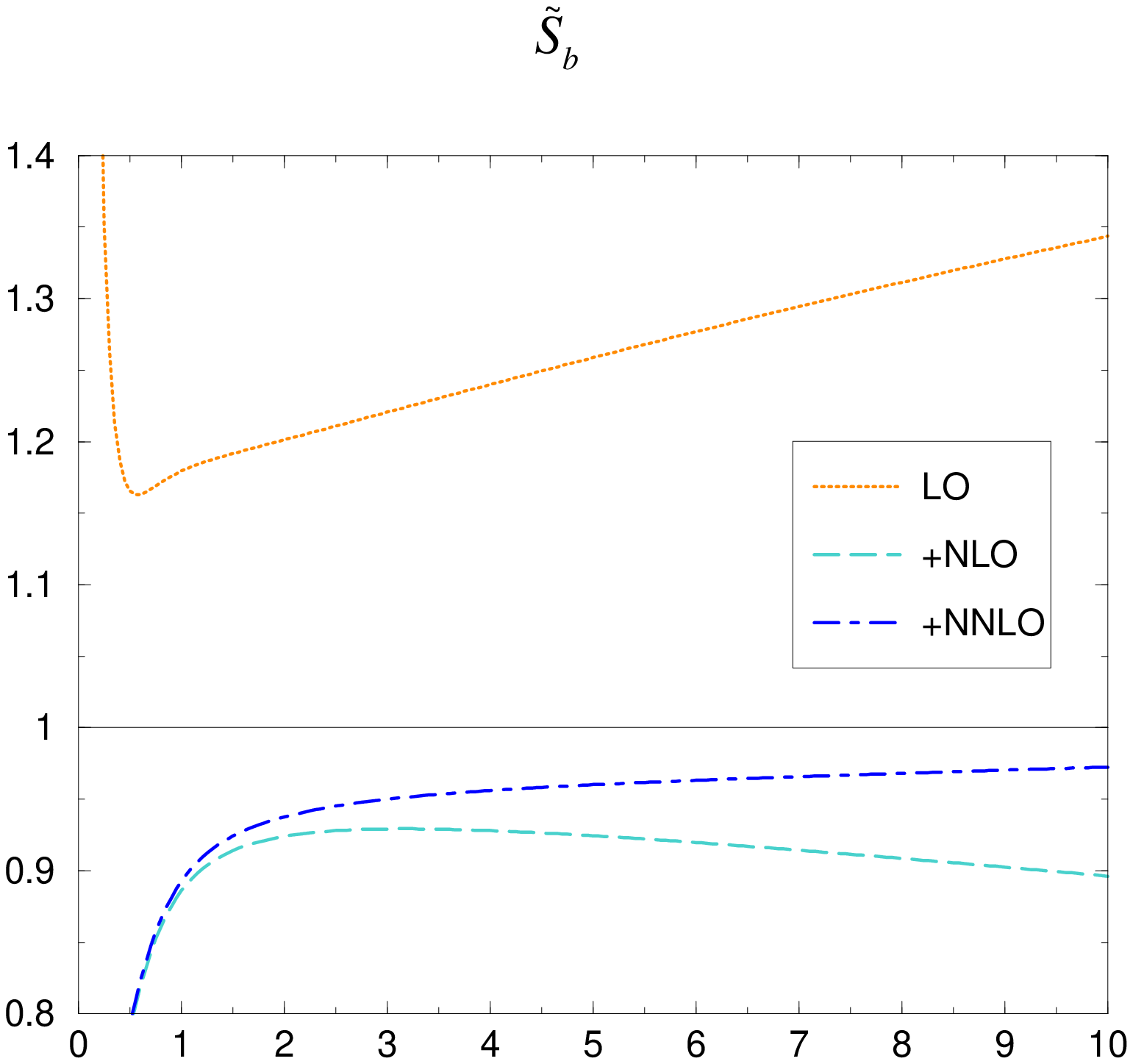, width=9cm}
\end{center}
\caption{The heavy baryon approximation of $\tilde S_b$. The partial 
sums are divided by the relativistic result, Eq. (\ref{3.24}).}
\label{fig:tPi_b}
\end{figure}

The quality of these heavy baryon approximations may be assessed in Figs. 
\ref{fig:Pi_x}--\ref{fig:tPi_b}, where the partial sums are divided by 
the relativistic result, Eqs. (\ref{3.20})--(\ref{3.24}).

\section{conclusions} \label{secVII}

In this work we have studied the main features, in configuration space, of 
a relativistic $O(q^4)$ expansion of the two-pion exchange nucleon-nucleon 
potential derived recently by ourselves \cite{HR}. Chiral symmetry provides 
a mathematical structure for the potential, that has to be fed with 
numerical values for $\mu$, $m$, $f_{\pi}$, $g_A$, and the LECs $c_i$ and 
$d_i$. The main source of uncertainty are the values of those LECs, which 
need to be extracted from $\pi N$ scattering data. 

The profile functions for the various non relativistic components 
of the potential were compared with two phenomenological versions 
produced by the Argonne group. One finds good agreement with their 
central scalar term, which 
dominates the $NN$ interaction. In all cases in which the signs of the 
Argonne potentials coincide, there is a qualitative agreement with our 
results. 

In order to check how empirical uncertainties can affect numerical results, 
we have studied the dynamical content of the 
potential in terms of families of diagrams associated with either the 
$[{\cal L}_{\pi}^{(2)}+{\cal L}_{N}^{(1)}]$ or 
$[{\cal L}_{N}^{(2)}+{\cal L}_{N}^{(3)}]$ pieces of the effective Lagrangian. 
In all but one cases, dynamics is clearly dominated by one of these 
interactions. In particular, $V_{LS}^{+}$, $V_{T}^{+}$, $V_{SS}^{+}$, and 
$V_{C}^{-}$ are dominated by $[{\cal L}_{\pi}^{(2)}+{\cal L}_{N}^{(1)}]$ 
and hence fixed by the values of $g_A$ and $f_{\pi}$ only. 
The components $V_{C}^{+}$, $V_{T}^{-}$, and $V_{SS}^{-}$, on the other 
hand, are dominated by $[{\cal L}_{N}^{(2)}+{\cal L}_{N}^{(3)}]$ and 
their numerical values may be affected by the less certain LECs $c_i$ 
and $d_i$. 

Most components of the potential are given as sums of $O(q^2)$, 
 $O(q^3)$, and $O(q^4)$ terms. The relative weights of these terms of the 
chiral series have been investigated and one finds good convergence at 
large distances. 
However, there are 
two cases, namely, $V_{C}^{+}$ and $V_{LS}^{-}$, 
where convergence is not evident in 
the region of physical interest. 
We intend to deal with this problem elsewhere. 

Finally, the relationship between relativistic and heavy baryon results 
has been discussed. On the purely conceptual side, the view seems to 
be well accepted nowadays that they cannot be fully equivalent. This is 
indeed the case and the  
numerical implications of this statement in configuration space were 
found to be of the order of 5\%. 

\section*{acknowledgments} 

The work of C.A.dR. was supported by Grant No. 97/6209-4 and 98/11578-1, 
and R.H., by Grant No. 99/00085-7, both from FAPESP (Funda\c c\~ao de 
Amparo \`a Pesquisa do Estado de S\~ao Paulo) Brazilian Agency.

\appendix 

\section*{} 

Several chiral calculations of the TPEP were produced in the last 
decade. As we pointed out in the introduction we expect, in the 
spirit of effective theories of QCD, that all these calculations 
should eventually converge to a single result. 

It is in this conceptual 
framework that we discuss here the relationship between the present work 
and its earlier versions, published between 1994 and 1997 
\cite{RR94,RR97}. 
In Ref.\cite{HR}, the $O(q^4)$ expansion of the TPEP was 
performed in three steps. In step 1, we derived 
full amplitudes, by using standard covariant techniques, 
to evaluate the diagrams of Fig. \ref{fig2}. At this stage, 
results were quite similar to those of Ref.\cite{RR97}, 
although not identical, 
as we discuss in the sequence. A handicap of the full amplitudes is 
that they involve several cancellations and do not exhibit 
chiral scales explicitly.
Therefore, in step 2 
we derived intermediate results, that show these scales, 
by just rewriting the full amplitudes with the help of exact relations 
among Feynman integrals. We subsequently neglected short distance terms 
and, in that region, full and intermediate 
results became no longer identical. The transmogrification 
of the potential was based on the following relations\footnote{For the 
complete details of the notation, please see Ref.\cite{HR}.}: 

\begin{eqnarray}
\mbox{Relation 01:}&& \bar{S}_{cc}^{(000)} =
\frac{1}{3}\left( 1-{\mbox{\boldmath $\nabla$}}^2/{4}\right) 
{S}_{cc}^{(000)}+\cdots\;,
\label{rel01}\\[2mm]
\mbox{Relation 02:}&& \bar{\bar{S}}_{cc}^{(000)}=
\frac{1}{15}\left(1-{\mbox{\boldmath $\nabla$}}^2/{4}\right)^2
{S}_{cc}^{(000)}+\cdots\;,
\label{rel02}\\[2mm]
\mbox{Relation 03:}&& \left[1-(\mu/m)^2\,
{\mbox{\boldmath $\nabla$}}^2/{4}\right]{S}_{sc}^{(001)}
= {S}_{cc}^{(000)} - \frac{\mu}{2m} \left( 1 -
{\mbox{\boldmath $\nabla$}}^2/{2}\right) {S}_{sc}^{(000)}+\cdots\;,
\label{rel03}\\[2mm]
\mbox{Relation 04:}&& \left[1-(\mu/m)^2\,
{\mbox{\boldmath $\nabla$}}^2/{4}\right]{S}_{sc}^{(002)}
+\bar{S}_{sc}^{(000)}
=-\frac{\mu}{2m}\left(1-{\mbox{\boldmath $\nabla$}}^2/{2}\right)
{S}_{sc}^{(001)}+\cdots\;,
\label{rel04}\\[2mm]
\mbox{Relation 05:}&& \bar{S}_{sc}^{(000)}=
\frac{1}{2}\left( 1-{\mbox{\boldmath $\nabla$}}^2/{4}\right)
{S}_{sc}^{(000)}+\frac{\mu}{4m}
\left( 1-{\mbox{\boldmath $\nabla$}}^2/{2}\right) {S}_{sc}^{(001)}+\cdots\;,
\label{rel05}\\[2mm]
\mbox{Relation 06:}&& \left[1-(\mu/m)^2\,
{\mbox{\boldmath $\nabla$}}^2/{4}\right]
{S}_{ss}^{(001)}={S}_{sc}^{(000)}
-\frac{\mu}{2m}\left(1-{\mbox{\boldmath $\nabla$}}^2/{2}\right)
{S}_{ss}^{(000)}+\cdots\;,
\label{rel06}\\[2mm]
\mbox{Relation 07:}&& \bar{S}_{ss}^{(000)}= - {S}_{sc}^{(001)}+\cdots\;,
\label{rel07}\\[2mm]
\mbox{Relation 08:}&& \left[1-(\mu/m)^2\,
{\mbox{\boldmath $\nabla$}}^2/{4}\right]{S}_{ss}^{(002)}
+\bar{S}_{ss}^{(000)}={S}_{sc}^{(001)}
-\frac{\mu}{2m}\left(1-{\mbox{\boldmath $\nabla$}}^2/{2}\right)
{S}_{ss}^{(001)}+\cdots\;,
\label{rel08}\\[2mm]
\mbox{Relation 09:}&& \left[1-(\mu/m)^2\,
{\mbox{\boldmath $\nabla$}}^2/{4}\right]^2{S}_{ss}^{(002)}
+\left[1-(\mu/m)^2\,
{\mbox{\boldmath $\nabla$}}^2/{4}\right]\bar{S}_{ss}^{(000)}=
\frac{\mu^2}{4m^2} \left( 1-{\mbox{\boldmath $\nabla$}}^2/{2}\right)^2 
\;{S}_{ss}^{(000)}
\nonumber\\[1mm]&&
+\left[1-(\mu/m)^2\,
{\mbox{\boldmath $\nabla$}}^2/{4}\right]{S}_{sc}^{(001)}
-\frac{\mu}{2m} \left( 1-{\mbox{\boldmath $\nabla$}}^2/{2}\right)
{S}_{sc}^{(000)}+\cdots\;,
\label{rel09}\\[2mm]
\mbox{Relation 10:}&& \bar{S}_{ss}^{(000)} =
\left(1-{\mbox{\boldmath $\nabla$}}^2/{4}\right) {S}_{ss}^{(000)}
+\frac{\mu}{2 m} \left(1-{\mbox{\boldmath $\nabla$}}^2/{2} \right) 
{S}_{ss}^{(001)}+\cdots\;,
\label{rel10}\\[2mm]
\mbox{Relation 11:}&& -{S}_{sc}^{(000)}=
+\frac{\mu}{2m}\left(1-{\mbox{\boldmath $\nabla$}}^2/{2}\right)
{S}_{reg}^{(000)}
-\frac{1}{\sqrt{1-(\mu/m)^2\,
{\mbox{\boldmath $\nabla$}}^2/{4}}}\,{S}_a+\cdots\;,
\label{rel11}\\[2mm]
\mbox{Relation 12:}&& \left[1-(\mu/m)^2\,
{\mbox{\boldmath $\nabla$}}^2/{4}\right]{S}_{reg}^{(002)}+
\bar{S}_{reg}^{(000)}={S}_{sc}^{(001)}+\cdots\;,
\label{rel12}\\[2mm]
\mbox{Relation 13:}&& \bar{S}_{reg}^{(000)}=-{S}_{sc}^{(001)}
+\frac{\mu}{2m}\left(1-{\mbox{\boldmath $\nabla$}}^2/{2}\right)
{S}_{reg}^{(010)}+\cdots\;,
\label{rel13}\\[2mm]
\mbox{Relation 14:}&& 
-\frac{\mu^2}{4m^2}\left(1-{\mbox{\boldmath $\nabla$}}^2/{2}\right)^2 
\;{S}_{reg}^{(000)}=
\nonumber\\[1mm]&&
\frac{\mu}{2m}\left(1-{\mbox{\boldmath $\nabla$}}^2/{2}\right)
{S}_{sc}^{(000)}
-\frac{\mu}{2m}\left(1-{\mbox{\boldmath $\nabla$}}^2/{2}\right)\,
\frac{1}{\sqrt{1-(\mu/m)^2\,
{\mbox{\boldmath $\nabla$}}^2/{4}}}\,{S}_{b}^{(000)}+\cdots\;,
\label{rel14}\\[2mm]
\mbox{Relation 15:}&& \bar{S}_{reg}^{(000)} =
\left( 1-{\mbox{\boldmath $\nabla$}}^2/{4}\right) {S}_{reg}^{(000)}
-\frac{\mu}{2 m} \left( 1-{\mbox{\boldmath $\nabla$}}^2/{2} \right) 
{S}_{reg}^{(010)}+\cdots\;.
\label{rel15}
\end{eqnarray}

In these expressions, the ellipses indicate short range terms, 
which have been neglected. In order to produce a feeling for the 
accuracy of these approximations, in table \ref{tab2} we display 
the quantity $\Delta_i(r)=|1-R_i(r)/L_i(r)|$, where 
$L_i(r)$ $R_i(r)$ are respectively the values of the left 
and right hand sides of relation $i$ at point $r$. 
Inspection of this table shows that, although discrepancies may be 
large at short distances, in all cases they remain below 1\% for 
$r\geq 1.5$ fm. 

\begin{table}[hb]
\begin{center}
\caption{Approximations made in relations among integrals.\label{tab2}}
\begin{tabular} {|c|c|c|c|c|}
\hline
\raisebox{0pt}[12pt][6pt]{}
 & $r=0.1$ fm & $r=0.5$ fm & $r=1.0$ fm & $r=1.5$ fm 
\\\hline
Rel. 01 & 0.000019 & 0.000000 & 0.000000 & 0.000000
\\ \hline
Rel. 02 & 0.000004 & 0.000000 & 0.000000 & 0.000000
\\ \hline
Rel. 03 & 0.004272 & 0.000001 & 0.000000 & 0.000000
\\ \hline
Rel. 04 & 0.000433 & 0.000000 & 0.000000 & 0.000000
\\ \hline
Rel. 05 & 0.002478 & 0.000001 & 0.000000 & 0.000000
\\ \hline
Rel. 06 & 0.056091 & 0.000002 & 0.000000 & 0.000000
\\ \hline
Rel. 07 & 0.000000 & 0.000000 & 0.000000 & 0.000000
\\ \hline
Rel. 08 & 0.005502 & 0.000010 & 0.000000 & 0.000000
\\ \hline
Rel. 09 & 0.271267 & 0.000347 & 0.000000 & 0.000000
\\ \hline
Rel. 10 & 0.034881 & 0.000009 & 0.000000 & 0.000000
\\ \hline
Rel. 11 & 0.958301 & 0.096190 & 0.000076 & 0.000312
\\ \hline
Rel. 12 & 0.000096 & 0.000030 & 0.000001 & 0.000000
\\ \hline
Rel. 13 & 0.855452 & 0.134455 & 0.006010 & 0.000168
\\ \hline
Rel. 14 & 1.007437 & 1.105614 & 0.066393 & 0.006432
\\ \hline
Rel. 15 & 1.554767 & 0.417655 & 0.018483 & 0.000513
\\ \hline
\end{tabular}
\end{center}
\end{table}

Finally, in step 3, we obtain the $O(q^4)$ 
expansion of the TPEP, given in section \ref{secIII}, 
by truncating the results of step 2 at that order. 

We now compare the results of this work with those from 
earlier versions. Our 1994 paper \cite{RR94} dealt with the 
evaluation of the diagrams given in family I of Fig.\ref{fig2}, which 
corresponds to the minimal realization of chiral symmetry in the TPEP. 
The main differences with our present results concern second order 
corrections, due to the way variable $W$, representing the total CM 
energy, was approximated in the planar box diagram. In 1994 paper we 
used $W=2m$, following Partovi and Lomon \cite{PL}. We no longer 
perform this crude approximation.

In our 1997 paper \cite{RR97} we have calculated the diagrams shown 
in family III of Fig.\ref{fig2} and results can be directly related 
with those of the present work, provided one establishes the connection 
between the two notations. For instance, the central isoscalar 
potential was formerly written as 

\begin{eqnarray}
\left. V^+_C\right|_{III}&=&
-\frac{\mu}{4\pi}\frac{3}{2}\left\{g^2\frac{\mu}{m}\alpha_{mn}^+
\left[2S_{B(2m,n)}+2S^V_{T(2m,n)}\right]
+\alpha^+_{k\ell}\alpha^+_{mn}S_{B(2k+2m,\ell+n)}\right. \nonumber \\ [3mm]
&+&g^2\frac{\mu}{m}\beta_{mn}^+\left[2S^V_{B(2m+1,n)}+2S^W_{T(2m+1,n)}\right]
+\alpha^+_{k\ell}\beta^+_{mn}S^V_{B(2k+2m+1,\ell+n)} \nonumber \\ [3mm]
&+&\left. \beta^+_{k\ell}\beta^+_{mn}S^W_{B(2k+2m+2,\ell+n)} + 
g^2\frac{\mu}{m}\beta_{mn}^+\,2S^g_{T(2m+1,n)}
+\beta^+_{k\ell}\beta^+_{mn}S^q_{B(2k+2m+2,\ell+n)}\right\}\;,
\label{eq01}
\end{eqnarray}

\noindent where $S$ are Feynman integrals from Ref.\cite{RR97} 
and $\alpha_{mn}^+$ and 
$\beta_{mn}^+$ are linear combinations of $\pi N$ subthreshold 
coefficients. In order to recast the old results 
in the form adopted in this work, one may use the relations 

\begin{eqnarray}
S_{B(0,n)} &=& \frac{(-1)^n}{4\pi}\left(1-{\mbox{\boldmath $\nabla$}}^2/4\right)^n\;S_{cc}^{000}
\label{eq04a} \\ [3mm]
S_{B(2,n)} &=& S^g_{B(0,n)}=S^V_{B(1,n)}= 
\frac{(-1)^n}{4\pi}\left(1-{\mbox{\boldmath $\nabla$}}^2/4\right)^n\;\bar{S}_{cc}^{000} 
\label{eq04b} \\ [3mm]
S_{B(4,n)} &=& 3S^g_{B(2,n)}=S^V_{B(3,n)}=(3/2)S^W_{B(2,n)}= 
\frac{(-1)^n}{4\pi}\left(1-{\mbox{\boldmath $\nabla$}}^2/4\right)^n\;\bar{\bar{S}}_{cc}^{000} 
\label{eq04c} \\ [3mm]
S^V_{T(0,n)} &=& \frac{(-1)^n}{4\pi}\left(1-{\mbox{\boldmath $\nabla$}}^2/4\right)^n\;
(-)S_{sc}^{001} 
\label{eq04d} \\ [3mm]
S^V_{T(2,n)} &=& S^W_{T(1,n)}+S^g_{T(1,n)}
\label{eq04e} \\ [3mm]
S^{W}_{T(1,n)} &=&
\frac{(-1)^n}{4\pi}\left(1-{\mbox{\boldmath $\nabla$}}^2/4\right)^n\;(-)\left[
S_{sc}^{003}+2\bar{S}_{sc}^{001}\right] 
\label{eq04f} \\ [3mm]
S^g_{T(1,n)} &=&-\frac{(-1)^n}{4\pi}\left(1-{\mbox{\boldmath $\nabla$}}^2/4\right)^n\;
(-)\bar{S}_{sc}^{001} 
\label{eq04g}
\end{eqnarray}

The parameters $\alpha^{+}_{mn}$ and $\beta^{+}_{mn}$ are related to 
subthreshold coefficients by 

\begin{eqnarray}
\alpha_{00}^+&=&\mu\left(\bar{d}_{00}^++4\mu^2\bar{d}_{01}^+
+16\mu^4\bar{d}_{02}^+\right) 
\label{eq5a} \\ [0.3cm]
\alpha_{01}^+&=&4\mu^3\left(\bar{d}_{01}^++8\mu^2\bar{d}_{02}^+\right) 
\label{eq5b} \\ [0.3cm]
\alpha_{02}^+&=&16\mu^5\left(\bar{d}_{02}^+\right) 
\label{eq5c} \\ [0.3cm]
\alpha_{10}^+&=&\mu^3\left[\bar{d}_{10}^+-{b}_{00}^+
+4\mu^2(\bar{d}_{11}^+-b_{01}^+)\right] 
\label{eq5d} \\ [0.3cm]
\alpha_{11}^+&=&4\mu^5\left(\bar{d}_{11}^+-{b}_{01}^+\right) 
\label{eq5e} \\ [0.3cm]
\beta_{00}^+&=&\mu^3\left({b}_{00}^++4\mu^2{b}_{01}^+
+16\mu^4{b}_{02}^+\right) 
\label{eq5f} \\ [0.3cm]
\beta_{01}^+&=&4\mu^5\left({b}_{01}^++8\mu^2{b}_{02}^+\right) 
\label{eq5g} \\ [0.3cm]
\beta_{02}^+&=&16\mu^7\left({b}_{02}^+\right) 
\label{eq5h}
\end{eqnarray}

\vspace{0.5cm}
Just as an example, using these rules in the case of the triangle 
contribution to Eq. (\ref{eq01}) and truncating at $O(q^4)$, we find 

\begin{eqnarray}
V_C^+\left. \right|_{III}^t&=&
-\frac{\mu}{4\pi}\frac{3}{2}g^2\frac{\mu}{m}2\left[\alpha_{0n}^+
S^V_{T(0,n)}+\alpha_{1n}^+S^V_{T(2,n)}+
\beta_{0n}^+\left(S^W_{T(1,n)}+S^g_{T(1,n)}\right)\right]
\label{eq06}
\\[5mm]&=&
+3[\mu]\left(\frac{g}{4\pi}\right)^2
\frac{\mu}{m}\left[\left(\bar{d}_{00}^++\bar{d}_{01}^+t
+\bar{d}_{02}^+t^2\right)S_{sc}^{001}
+\left(\bar{d}_{10}^++\bar{d}_{11}^+t\right)
\left(S_{sc}^{003}+3\bar{S}_{sc}^{001}\right)\right]\,.
\label{eq07}
\end{eqnarray}

We have recently checked explicitly all the results of our 1997 paper 
and found out that they are equivalent with those of the present work 
if we make the approximation $W=2m$.

There are still two important sources of differences between these 
two sets of results. The first one is due to the fact that those of the 
earlier work were not truncated at a given order. The second one is that 
it did not include explicitly the two-loop diagrams, as we do now. 
In 1997 these effects were hidden within the $\pi N$ sub-threshold 
coefficients and were therefore double counted. Even if these effects 
are numerically small, as we discussed in section \ref{secIV} of this 
work, this represents a rather important conceptual difference between 
both calculations. As two-loop contributions only arise at 
$O(q^4)$, the potential produced in 1997 would be numerically 
identical with the present one for distances larger than 1.5 fm if 
both of them were truncated at $O(q^3)$. 


\end{document}